\documentclass[pra,aps,twocolumn,amsfonts,showpacs]{revtex4}
\usepackage{epsfig,amsmath}
\def\qed{\leavevmode\unskip\penalty9999 \hbox{}\nobreak\hfill
     \quad\hbox{\leavevmode  \hbox to.77778em{%
               \hfil\vrule   \vbox to.675em%
               {\hrule width.6em\vfil\hrule}\vrule\hfil}}
     \par\vskip3pt}
\def\ibb #1{\leavevmode\hbox{\kern.3em\vrule
     height 1.5ex depth -.1ex width .4pt\kern-.3em\rm#1}}

\newcommand{\be}{\begin{equation}}
\newcommand{\ee}{\end{equation}}
\newcommand{\bea}{\begin{eqnarray}}
\newcommand{\eea}{\end{eqnarray}}
\newcommand{\la}{\langle}
\newcommand{\ra}{\rangle}

\def\bma{\begin{mathletters}}
\def\ema{\end{mathletters}}
\newcommand{\one}{\mbox{$1 \hspace{-1.0mm}  {\bf l}$}}

\def\C{\hbox{$\mit I$\kern-.7em$\mit C$}}
\newcommand{\tr}{{\rm tr}}

\newcommand{\ket}[1]{ | \, #1  \rangle}

\begin{document}
\title{Localizable Entanglement}
\author{M. Popp$^{1}$, F. Verstraete$^{1,2}$, M. A.  Martin-Delgado$^{1,3}$  and  J. I.  Cirac$^{1}$ }
\affiliation{$^{1}$Max-Planck-Institut f{\"u}r Quantenoptik,
         85748 Garching, Germany\\
         $^{2}$Institute for Quantum Information, Caltech, Pasadena, CA 91125  USA\\
$^{3}$Departamento de F\'isica Te\'orica I, Universidad Complutense de Madrid,
E-28040, Spain}
\date{\today}
\begin{abstract}
We consider systems of interacting spins and study the
entanglement that can be \emph{localized}, on average,  between
two separated spins by performing local measurements on the
remaining spins. This concept of \emph{Localizable Entanglement}
(LE)  leads naturally to notions  like entanglement length and
entanglement fluctuations. For both spin-1/2 and spin-1 systems we
prove that the LE of a pure quantum state can be lower bounded by
connected correlation functions. We further propose a scheme,
based on matrix-product states and the Monte Carlo method, to
efficiently calculate the LE for quantum states of a large number
of spins. The virtues of LE are illustrated for various spin
models. In particular, characteristic features of a quantum phase
transition such as a diverging entanglement length can be
observed. We also give examples for pure quantum states
 exhibiting a diverging entanglement length but finite correlation length. We have numerical evidence that the ground state of
the antiferromagnetic spin-1 Heisenberg chain can serve as a perfect quantum
channel. Furthermore, we apply the  numerical method  to mixed states and study the  entanglement as a function of temperature.

\end{abstract}
\date{\today}
\pacs{03.67.Mn, 03.65.Ud, 75.10.Pq, 73.43.Nq }
\maketitle

\section{Introduction}
The creation and distribution of entangled states plays a central role in quantum
information, because it is the key ingredient for performing certain
quantum information tasks, like teleportation \cite{teleportation} or quantum computation. In this
respect, multiparticle quantum states can be considered as entanglement resources,
 naturally appearing in many physical systems. On the other hand, it is believed
that the study of multipartite entanglement might prove fruitful
in other fields of physics, like condensed matter, e.g. for
understanding the complex physics of strongly correlated states
\cite{P00,ZW02}. In particular it has been shown
\cite{O02,N02,V02,Vedral01,Ortiz04,Tognetti04} that the ground
state entanglement of various spin systems may exhibit
characteristic features at a quantum phase transition. Hence it is
desirable to find ways of characterizing and quantifying
entanglement in multipartite systems.

In \cite{VPC04}  an entanglement measure,  called the \emph{Localizable
Entanglement} (LE), has been defined. It quantifies the bipartite entanglement
contained in a multipartite system. The concept of LE allows one to define the notion
of entanglement length, which characterizes the typical distance up to which
bipartite entanglement can be localized in the system.  Moreover, it has been
shown  for general  pure qubit states that  the LE can  always be lower bounded by
connected correlation functions \cite{VPC04}.  Hence quantum phase transitions,
characterized by a diverging  correlation length, are equivalently detected by a
diverging entanglement length. In fact, the concept of LE has already proven useful in several studies of entanglement properties of spin systems \cite{VPC04, VMC04,
JK03,LEcontr,R04}. Indeed, for ground states of many spin-1/2 models one finds
that correlations and entanglement (as given by LE) typically exhibit  the same
qualitative behavior. In gapped \mbox{spin-1} systems,  which exhibit finite
correlation lengths \cite{Hastings}, however, a ground state has been found for which the
entanglement length is infinite \cite{VMC04}. This example also shows that the
concept of LE can serve to detect hidden order in certain states.

The LE is the maximum entanglement that can be localized, on average, between two parties of a multipartite system, by performing local measurements on the other parties. Hence, LE is defined in an operational way and has a clear physical meaning. For instance it can be used as a figure of merit to characterize the performance of quantum repeaters \cite{Br98}. Note that, in the context of LE, particles are not traced out but measured.
This is in contrast  to earlier approaches, where the concurrence of the reduced density operator of two separated spins in a spin chain has been calculated \mbox{(see e.g. \cite{O02, N02})}. Although the concurrence exhibited   characteristic features at a  quantum phase transition, it does not  detect long range quantum correlations.

The fundamental difference between tracing and measuring  can  be illustrated with two simple examples:
For both the GHZ state \cite{GHZ} and the cluster state \cite{Br01} it can readily be checked that the reduced density operator of any two qubits contains no entanglement at all. On the other hand one can find a local measurement basis, such that the LE is maximal.
We further note that in the case of the GHZ-state the entanglement properties
could also have been revealed by studying the connected version of the two-point
correlation function. Indeed, the GHZ-state is one of many examples \cite{VPC04},
for which correlations can  be identified with quantum correlations, i.e.
entanglement. However, this intimate connection does not hold true for all pure
quantum states (as  shown in \cite{VMC04}). For instance, in the case of  the
cluster state the LE  is maximal, whereas the connected correlations are all zero.

The intention of this paper is to  provide a framework for using LE as an measure for localizable bipartite entanglement of multipartite quantum systems. Starting from the original definition  \cite{VPC04}, we introduce several variants of LE,
 and establish basic relations between them. Apart from the notion of entanglement length , we also give a meaning to the notion of \emph{entanglement fluctuations} in terms of LE.  Moreover we  generalize the earlier result, that the LE is lower bounded by connected two-particle  correlation functions, to pure qutrit states, e.g. ground states of spin-1 systems.
Apart from these analytical findings we  present a method for the numerical computation of the LE for arbitrary one
dimensional spin systems. It allows us to efficiently simulate chains  with even
more than $100$ sites and also works  for finite temperatures. We apply  this
method for the study of LE present in ground states of various spins models. We
find that the  entanglement fluctuations as well as the LE exhibit characteristic
features at a quantum phase transition. For instance,  we observe a discontinuity in the first derivative of the LE at a Kosterlitz-Thouless transition \cite{Sachdev}. These findings can be understood as a direct consequence  of the numerical observation, that for spin-1/2 systems the LE and the maximal connected
correlation function are typically equal to each other.  In the case of the spin-1
Heisenberg antiferromagnet, however,  we rather observe the  opposite behavior,
namely an infinite entanglement length but finite correlation length. In this context we  also comment on a possible connection between the presence of  hidden order in the state and the existence of long range entanglement. In particular, we present examples showing that, in general, such a connection does not exist.  As a  further application of our numerical method we
study  the effect of  finite temperature on the LE. For the AKLT model
\cite{AKLT} we find  that the entanglement length  increases exponentially with
the inverse temperature.

The paper is organized as follows: We start by giving a formal definition of LE
and derive quantities like entanglement length and entanglement fluctuations. We
further connect the concept of LE with the idea of quantum repeaters. In Sect. III
we recapitulate the earlier result for the lower bound of LE in terms of connected
correlation functions and generalize it to pure qutrit systems. Next, we   present
in Sect. IV a numerical method to compute the LE for spin chains. The numerical
scheme  is based on  the matrix-product state (MPS) \cite{Fannes, RomerPRL} representation of ground
states and  the Monte Carlo (MC) \cite{Metro53} method. In Sect. V this method is applied to
calculate the LE and the entanglement fluctuations for ground states of various
standard spin-1/2 chains. In Sect. VI we study the LE of gapped spin-1 models. In
particular, we present numerical calculations for the ground state of the
Heisenberg antiferromagnet, showing that the entanglement length diverges.
Furthermore, we comment on a possible connection between a diverging entanglement
length and hidden order in the system.  In   Sect. VII  we demonstrate that our
numerical method can also be used to calculate the LE of mixed states. As an
example we compute the entanglement length as  a function of the temperature for
the AKLT model. In Appendix A we present an extended version of the proof in
\cite{VPC04}, leading to the lower bound of LE. We connect in Appendix
B the entanglement of pure two qubit and two qutrit states to the maximum
connected correlation function of those states. Finally, in Appendix C, we show how
to calculate the LE (and the string order parameter) analytically for pure states, represented  by  MPS with qubit bonds.

\section{Definition and basic properties of Localizable Entanglement}
We consider a multipartite system composed of $N$ particles. With each particle we associate a finite dimensional Hilbert space. For simplicity we refer in the following to the particles as spins.
\subsection{Definition}
The Localizable Entanglement (LE)
of a multi-spin state $\rho$ is defined as the
maximal amount of entanglement that can be created (i.e. localized), on average,
between two spins at positions $i$ and $j$ by performing local measurements on the other spins.
More specifically, every measurement $\mathcal{M} $ specifies a state ensemble
${\mathcal{E}_\mathcal{M}} := \{p_s ,  \rho_s^{ij} \} $.
Here $p_s$ denotes the
probability to obtain the (normalized) two-spin state $\rho_s^{ij}$ for the outcome $\{ s \}$ of the measurements on the  $N-2$ remaining spins. The average entanglement for a specific  $\mathcal{M}$ is then given by:
\be \label{LEav}
 \overline{L}^{\mathcal{M},E}_{i,j}(\rho):=
\sum_s p_s \ E(\rho_s^{ij}) ,
\ee
where $E(\rho_s^{ij})$ is the entanglement
of $\rho_s^{ij}$. Suitable entanglement measures will be discussed later in this section.
The Localizable Entanglement is defined as the largest possible average entanglement:
\be \label{LE}
 L^{\mathcal{C},E}_{i,j}(\rho):= \sup_{\mathcal{M} \in \mathcal{C}}
\sum_s p_s \ E(\rho_s^{ij}), \ee with $\mathcal{C}$ denoting the class of allowed
measurements. We call the measurement $\mathcal{M}$ which maximizes the average
entanglement the \emph{optimal basis}. It is important to note that the only
restriction on $\mathcal{M}$ is that the measurements are performed locally i.e.
on individual spins. Apart from that, the measurment basis is arbitrary and can
also vary from site to site. We  distinguish three classes $\mathcal{C}$ of
measurements: projective von-Neumann measurements (PM), those corresponding to
positive operator-valued measures (POVM), and general local measurements that
allow also for classical communication of measurement results (LOCC). In terms of
LE the following relationship between these classes holds \cite{comLOCC}: \be
\label{LEclasses}
 L^{{\rm{PM}},E}_{i,j}(\rho) \leq  L^{{ \rm{POVM}},E}_{i,j}(\rho) \leq
L^{{\small \rm{LOCC}},E}_{i,j}(\rho).
\ee
In this paper we will be mainly concerned with projective measurements. To simplify the notation we omit in this case the superscript PM.

The definition (\ref{LE}) still leaves open the choice of the entanglement measure $E$ for the states $\{ \rho_s^{ij} \}$. Suitable measures depend on aspects  like the dimensionality of the spins and the purity of the state. In the following we specify for the cases of both pure and mixed spin-1/2 and spin-1 systems appropriate entanglement measures used in this article.

For  \emph{pure} multipartite states $\rho$, the states  $ \{ \rho_s^{ij} \}$ after the measurements  are also pure, and then there exists, in principle, a special entanglement measure
\cite{Bennett96}, i.e. the entropy of entanglement.
For a pure bipartite state, $\rho_{ij}=|\psi \ra \la \psi |$, the  entropy of entanglement $E_E$  is defined as the von Neuman entropy of the reduced density operator \mbox{$\rho_i=\tr_j (\rho_{ij})$}:
\be \label{EEdef}
E_E(\rho_{ij})=-\tr ( \rho_i \log_2 \rho_i) .
\ee
In the case of a pure two qubit state, $|\psi\rangle$, it can be shown that the  $E_E$  is a convex,
monotonously increasing function, $E_E=f(C)$, of the concurrence  \cite{Woo98}. The conurrence  $C(\psi)$ for pure states and the convex function $f$ are defined as:
\begin{eqnarray}
C(\psi)&:=&|\langle\psi^*|\sigma_y\otimes\sigma_y|\psi\rangle| \label{DefC} , \\
f(C)&:=&H\left(\frac{1+\sqrt{1-C^2}}{2}\right) \label{convexf} , \\
H(x)&:=&-x\log_2x-(1-x)\log_2(1-x) . \end{eqnarray}
 Here, $|\psi^*\rangle$ denotes the complex conjugate of $|\psi\rangle$ in the
standard basis and $H(x)$ is the Shannon entropy. In this article we will
typically use the concurrence to measure the entanglement of two qubits, because
it can be simply related  to connected correlation functions (see \mbox{Appendix
B}). However, due to the convexity of the function $f(C)$, the LE as measured by
the concurrence, $L_{ij}^C$, yields lower and upper bounds for the LE as measured
by the entropy of entanglement, $L_{ij}^{E_E}$:
 \be f\left( L_{ij}^C \right)\leq L_{ij}^{E_E} \leq   L_{ij}^C .
\ee
Therefore the qualitative behavior of these two variants of LE will be very  similar.\\
In the case of  \emph{mixed} states  the entanglement can be  calculated
using, e.g., the entanglement of formation $E_F$ \cite{EoF} or the negativity \cite{Negativity}.
For two qubits an explicit formula for the $E_F$ in terms of the concurrence $C$
 exists, which reduces to (\ref{DefC}) in the case of pure states.
The negativity $N$ can  in principle be calculated for any spin dimension.

\subsection{Entanglement length and fluctuations}
 In the field of strongly correlated systems and more
specifically in the study of quantum phase transitions, the correlation length, $\xi_C$,  is of great importance.
 The concept of LE readily lends itself to
define the related \emph{entanglement length}, $\xi_{E}$, as the typical length
scale at which it is possible to create Bell states by doing local measurements on
the other spins:
\be
\xi_{E}^{-1}:=\lim_{n\rightarrow\infty}\left(\frac{-\ln
L^{E}_{i,i+n}}{n}\right) .
\ee
The entanglement length is finite iff the LE \mbox{$L^{E}_{i,i+n} \rightarrow \exp(-n/\xi_E)$} for $n \rightarrow \infty$, and the entanglement length $\xi_E$ is defined as the
constant in the exponent in the limit of an infinite system (see also Aharonov
\cite{Ahar00}).

Let us now  have a closer look at the statistical nature of LE, as it is defined
as an average over  all possible measurement outcomes (see (\ref{LE})). For practical purposes one
can only control the measurement basis but not a specific outcome. Therefore it
would be useful to have an estimate of how much the entanglement of a particular
measurement outcome deviates from the mean value as given by the LE. This
information is contained in the variance of the entanglement remaining after measurements. We can thus  define  the   notion of
\emph{entanglement fluctuations}:
\be \label{def-fluc} (\delta L^{\mathcal{M},E}_{i,j})^2:= \left( \sum_s p_s \
{E(\rho^{ij}_s)}^2 \right) -{L^{\mathcal{M},E}_{i,j}}^2.
\ee
The entanglement fluctuations can be defined for any measurement $\mathcal{M}$. Typically we choose for $\mathcal{M}$ the optimal basis, which maximizes the  average entanglement. In this case we drop the index $\mathcal{M}$ in (\ref{def-fluc}).

The study of both the entanglement length and the  entanglement fluctuations could provide further inside in the complex physics of  quantum phase transitions by revealing characteristic features at the quantum critical point. Examples for this are presented in Sect. V.

\subsection{Connection to quantum repeaters}
So far we have given a purely mathematical definition of LE
(\ref{LE}). However, it is evident that the LE is defined in an
operational way that can directly be implemented on certain
physical systems.  In addition the concept of LE may also play an
essential role in some interesting applications of quantum
information theory. To be more precise, LE can serve as a figure
of merit for the ``performance'' of certain kinds of quantum
repeaters (QR).

\begin{figure}[h]
\begin{center}
      \epsfig{file=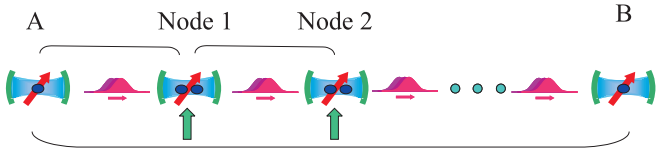,angle=0,width=\linewidth}
\end{center}
\caption{Illustration of the quantum repeater scheme for trapped atoms connected by optical fibres. Qubits are represented by the internal states of the atoms. Applying laser beams the internal states of  atoms in adjacent cavities become entangled via the transmission of photonic states. Collective measurements on the nodes (indicated by arrows) followed by purification lead to the generation of a nearly perfectly entangled pair of qubits between A and B.}
       \label{fig-QR}
\end{figure}

Many tasks in quantum information processing require long-distance
quantum communication. This means quantum states have to be
transmitted with high communication fidelity via a quantum channel
between two distant parties, Alice and Bob. Since quantum
transport is also possible via teleportation \cite{teleportation}
this problem is equivalent to establishing nearly perfect
entanglement between two distant nodes. All realistic schemes for
quantum communication are presently based on the use of photonic
channels. However, the degree of entanglement generated between
distant sites typically  decreases exponentially with the length
of the connecting physical channel, due to light absorption and
other channel noise. To overcome this obstacle the concept of
quantum repeaters has been introduced \cite{Br98}. The central
idea is to divide the channel into segments and to include
additional  nodes.  Entanglement between adjacent nodes can be
extended to larger distances using entanglement swapping followed
by purification. After several rounds one obtains a pair of almost
maximally entangled nodes, shared by Alice and Bob, that can be
used for perfect quantum transport via teleportation. A possible
physical realization of the QR using trapped atoms  is sketched in
Fig. \ref{fig-QR}  \cite{QRatoms}.

Let us now discuss how a QR setup can be characterized by the LE.
First of all, let us identify the particles sitting at different
nodes by spins. It is important to note that, by combining several
spins to a larger Hilbert space of dimension $d$,  the operations
required for purification and entanglement swapping on this  set
of spins can be interpreted as local operations on a single spin
of dimension $d$. Thus the QR can be treated as a system of
interacting spins being in a state $\rho$. In order to assess and
quantify the usefulness of such a setup as a QR one has to compute
the following figure of merit: What is the maximum amount of
entanglement that can be generated between the two end spins by
performing local operations on the intermediate spins? But this
number is nothing else than the LE. The question, which variant of
LE (\ref{LEclasses}) should be used, depends on the class of
available local operations (PM, POVM or LOCC). Typically classical
communication is allowed so that $L_{ij}^{\rm LOCC}$ has to be
taken as figure of merit. However, not every measurement might be
physically realizable. Therefore  the LE will in general  give an
upper bound for the performance of a given QR setup.

\section{Bounds on LE}
Due to its variational definition, the LE is very difficult to calculate in
general. Moreover,  typically  one does not have an explicit parameterization of
the state under interest, but just information about the classical one- and
two-particle correlation functions (which allows one to parameterize completely
the two-spin reduced density operator  $\rho_{ij}$). It would therefore be
interesting to derive tight upper and lower bounds to the LE solely based on this
information.
\subsection{Upper bound}
The upper bound can readily be obtained using the
concept of entanglement of assistance (EoA) \cite{DiV98}. The EoA depends only on the reduced density operator $\rho_{ij}$  for spins $i$ and $j$ and can be defined for any spin dimension. In the  case of
a pair of qubits an explicit formula for EoA can be derived \cite{V01}. Given  $\rho_{ij}$ and a square root $X$,
$\rho_{ij}=XX^\dagger$, then the EoA as measured by the concurrence reads: \be
\label{EoA} E_A^{i,j}(\rho_{ij}):=\tr|X^T(\sigma_y^i\otimes\sigma_y^j)X|  \ ,
\ee
with $|A |=\sqrt{A^\dagger A}$. Hence $E_A^{i,j}(\rho_{ij})=\sum_{k=1}^4 \sigma_k$, where $\sigma_k$ are the singular values of the matrix $X^T(\sigma_y^i\otimes\sigma_y^j)X$.
Note that for pure four qubit states a variant of EoA with local measurements was considered in \cite{V01}.
\subsection{Lower bound}
First,  from the definition of \mbox{LE (\ref{LE})} it follows that any specific measurement,  e.g. in the computational basis, trivially provides a lower bound on the LE. More interestingly,
 it has been proven in \cite{VPC04} that for general \emph{pure} qubit states the LE can be lower  bounded by connected correlation functions. In the following we recapitulate the central results leading to this bound for spin-1/2 systems and present some extensions. Next we show to what extent these findings can be generalized to higher dimensional spin systems.
\subsubsection{Spin-1/2 systems}
The basic idea is to establish a connection between the  LE and connected
correlation functions of the form: {\small \be \label{Qab1}
 Q_{AB}^{ij}= \tr[\rho ( \ S^i_A \otimes S^j_B)]-\tr[\rho ( \
S^i_A \otimes \one)] \tr[\rho  (\one \otimes S^j_B)]. \ee}
\hspace*{-2mm} For qubits the operators $S_A, S_B$ can be parameterized by directions $\vec{a}, \vec b$, representing unit vectors in a 3D real space:  $S_A=\vec a \cdot \vec{\sigma}, S_B=\vec b \cdot \vec \sigma$ with $ \vec{\sigma}=(\sigma_x, \sigma_y,\sigma_z)$. \\
We start out by quoting the central result of \cite{VPC04}:\\
\emph{(1.i) Given a (pure or mixed) state of $N$ qubits with connected correlation
function $Q^{ij}_{A B}$ between the spins $i$ and $j$ and directions $\vec{a},
\vec{b}$, then there always exists a basis  in which one can locally measure the
other spins such that this correlation does not decrease, on average.}
\\
An extended version of the proof is presented in Appendix A. There we also show that this result can be generalized to a setup, where the spins $i$ and $j$ can be of any dimension, but the remaining spins (on which the measurements  are performed) are still qubits. In a spin-1/2 system such a situation can arise, for example, when considering correlations between two blocks of spins.

Next, we relate correlations with entanglement. We note that after the measurement process and for an initially pure state we end up with a pure state of two qubits.
For such a state we have proven the following result \cite{VPC04} (see also Appendix B): \\
\emph{(1.ii) The entanglement of a pure two qubit state $|\psi_{ij} \ra$ as measured by the concurrence is equal to the maximal correlation function:}
\be \label{CQqubit}
C(\psi_{ij})= \max_{\vec{a}, \vec{b}} | \ Q_{AB}^{ij}(\psi_{ij}) |.
\ee
Combining \emph{(1.i)} and \emph{(1.ii)} we know that for a given pure multi qubit state $| \psi \ra $ and directions $\vec{a}, \vec{b}$ there always exists a measurement $\mathcal{M}$ such that:
\be \label{QLrel}
Q_{AB}^{ij}(\psi)\leq \sum_s p_s Q_{AB}^{ij}(\psi_s) \leq \sum_s p_s C(\psi_s)
.
\ee
The term on the very right is equal to the average entanglement as measured by the concurrence $ L_{ij}^{\mathcal{M},C}$, which trivially is a lower bound to the LE as defined by $ L_{ij}^{C}$. Since the directions  $\vec{a}, \vec{b}$ can be chosen arbitrarily, relation (\ref{QLrel}) holds in particular for directions maximizing $Q_{AB}^{ij}(\psi)$. Hence we can establish the desired lower bound on LE \cite{VPC04}:\\
\emph{(1.iii) Given a pure state $| \psi \ra$ of N qubits, then the LE
 as measured by the concurrence is larger or equal than the
maximal correlation:}
\be \label{LElb}
L^C_{i,j}(\psi ) \geq \max_{\vec{a}, \vec{b}} |Q_{A B}^{ij}(\psi)|  .
\ee
Making use of the basic properties of LE, presented in the previous section, we can immediately derive analogous bounds for some other variants of LE; for example,
\bea
L^{E_E}_{i,j}(\psi )& \geq& f(\max_{\vec{a}, \vec{b}} |Q_{A B}^{ij}(\psi)|), \label{LEElb}\\
L^{\rm{POVM},E_E}_{i,j}(\psi )& \geq& f(\max_{\vec{a}, \vec{b}} |Q_{A B}^{ij}(\psi)|), \label{LPOVMlb}
\eea
with $f$ being the convex function defined in (\ref{convexf}). We will see below that  relation (\ref{LPOVMlb}) can be generalized to spin-1  systems.
\subsubsection{Higher dimensional spin systems}
We now try to extend the previous findings beyond spin-1/2 systems. First, we look for a generalized version of statement \emph{(1.i)}. Unfortunately the techniques used in the proof for qubits seem to fail already for qutrits. Nevertheless, a generalization is still possible by changing a little bit the perspective. For this we embed a spin-$S$ in a higher dimensional Hilbert space,  being composed of
 $n\geq \log_2(2S+1)$ \emph{virtual} qubits. Let us denote the  $(2S+1) \times 2^n$ matrix governing this transformation by $P$. \\
 In the case $2S+1=2^n$ the embedding is trivial and the situation becomes equivalent to the qubit case. Thus the result  \emph{(1.i)} can immediately be generalized, because  local measurements on the virtual qubit systems can be chosen such that  \emph{(1.i)} holds.
  \\
In the case $2S+1< 2^n$  a similiar argument applies if we allow for POVM measurements on the spin-$S$ system.  To be more precise, let us consider a mixed state  $\rho$ of three spin-$S$ particles. The spin on which the measurement is performed (let us denote it with the \mbox{index $3$}) is embedded in a $2^n$ dimensional system. The embedded state is then given by the transformation: $\rho'=(\one_{12}\otimes P_3^\dagger) \rho (\one_{12}\otimes P_3)$. In the Hilbert space of the $n$ virtual qubits one always finds local projective measurements
$\{ M_{\alpha_1 \ldots \alpha_n} \} = \{ |\alpha_1 \ra  \la \alpha_1| \otimes \ldots \otimes |\alpha_n \ra  \la \alpha_n| \}$  such that (the generalized version of) \emph{(1.i)} holds for the state $\rho'$. In  terms of the original state $\rho$ this measurement in the $2^n$ dimensional space corresponds to a POVM measurement $\{ P M_{\alpha_1 \ldots \alpha_n} P^\dagger \} $ on a  spin-S system, because: $\sum_{\alpha_1 \ldots \alpha_n}  P M_{\alpha_1 \ldots \alpha_n} P^\dagger=P \one_{2^n \times 2^n} P^\dagger=\one_{(2S+1) \times (2S+1)}$.

 Thus one can  generalize the result \emph{(1.i)}   to arbitrary spin dimensions in the following way:\\
\emph{(2.i) Given an arbitrary multi-spin  state with connected correlation
function $Q^{ij}_{A B}$ between  spins $i$ and $j$ for arbitrary operators $S_A,
S_B$, then there always exists a local POVM measurement on  the other spins such
that this correlation does not decrease, on average.}
\\
 We  note that the
lower bound in \emph{(2.i)} can already be reached by applying \emph{local} measurements on the
 virtual qubit system. Performing joint measurements (e.g. Bell measurements on
pairs of qubits as shown in \cite{VMC04}) can lead to a considerable enhancement
of the average correlations. If in addition $2^n-(2S+1)$  joint measurements can be chosen such that they are orthogonal to the projector $P$, the resulting measurement on the spin-S system corresponds to a projective von-Neumann measurement.

The most difficult part is to establish a connection between
correlations and entanglement for pure two spin states in analogy
of \emph{(1.ii)}. In the case of qubits we made explicitly use of
the fact that the group $SU(2)$ is  the covering group of
 $SO(3)$. Moreover the concurrence served as an entanglement measure, which was easy to handle. For higher spin dimensions we  refer to the entropy of
entanglement $E_E$ (\ref{EEdef}) as a suitable entanglement measure for pure bipartite states. In the special case of qutrits we were able to show the following relation (Appendix B):\\
\emph{(2.ii) The entanglement of a pure two qutrit state $|\psi_{ij} \ra$ as measured by the entropy of entanglement can be lower bounded by:}
\be \label{EEQqutrit}
E_E(\psi_{ij})\geq f(\max_{A,B} | \ Q_{AB}^{ij}(\psi_{ij}) |),
\ee
where $f$ is the convex function (\ref{convexf}) and  $S_A, S_B$ in $Q_{AB}^{ij}$ (\ref{Qab1}) are operators, whose eigenvalues lie in the interval $[-1;1]$.
Combining again \emph{(2.i)} and \emph{(2.ii)} we
 can formulate a bound on LE  for spin-1 systems:\\
\emph{ (2.iii) Given a pure state $ | \psi \ra$ of $N$ qutrits, then the LE  as
measured by the entropy of entanglement and
 which allows for \rm{POVM's}, is lower bounded by the maximum connected correlation function in the following way:}
\be L^{\rm{POVM},E_E}_{ij}(\psi ) \geq f(\max_{A, B} |Q_{A B}^{ij}(\psi)| ).
\ee

In summary, we have shown for pure qubit and qutrit states that connected
correlation functions  provide a lower bound on LE. This bound allows for two
intriguing limiting case: (i) Entanglement and correlations may exhibit similiar
behavior. (ii)  Spins may be maximally entangled although they are uncorrelated in
the classical sense. In the forthcoming sections we will present examples for both
scenarios.

\section{Computation of LE based on matrix product states and the Monte Carlo method}
In this section we propose different techniques to calculate the LE numerically for ground states of 1D spin systems.
Let us consider a chain of $N$ spins of dimension $d=2S+1$. The ground state of
the system can be determined exactly by diagonalization of the Hamiltonian. However, the
 state  is characterized by an exponential amount of  parameters as a
function of $N$, thus  limiting the exact treatment of the problem to relatively
small system sizes.

As an alternative approach we can start out with an approximation of the exact ground state in terms of the
 the so-called
matrix product states (MPS)  \cite{RomerPRL,Fannes}:
\begin{equation}
  \label{MPS}
 |\psi_{\mathrm{MP}}\rangle = \sum_{s_1,\ldots,s_N=1}^d {\rm Tr}
 (A^{s_1}_1 \ldots A^{s_N}_N) |s_1,\ldots,s_N\rangle.
\end{equation}
Here the state is described by $N$  matrices $ A^{s_i}_i$ of dimension $D$.
We note that the MPS (\ref{MPS}) is written in the computational basis and  accounts for periodic boundary
conditions (PBC). It has been shown \cite{RomerPRL,D97} that MPS  appear naturally in the context of  the
density-matrix renormalization group (DMRG) method \cite{W92}.
Assuming we are able to  calculate the ground state in MPS form (\ref{MPS}), let us now present a scheme to compute the LE from that. For translationally invariant systems
  it is sufficient  to consider the LE between the spins $1$ and $j=1+n$. The (pure)  normalized state of
these two spins, after local  measurements on the remaining ones have been performed, is conditioned on
the measurement outcomes denoted by the (N-2)-tuple $\{s\}:=\{s_2 \ldots s_{j-1}
s_{j+1} \ldots s_N\}$ and proportional to
 \be \label{phis} |\phi_{\{s\}} \ra = \la
\{s\}|\psi_{\mathrm{MP}}\rangle =  \sum_{s_1,s_j=1}^d {\rm Tr}
 (A^{s_1}_1 \ldots A^{s_N}_N) |s_1\ra |s_j \rangle.
\ee
Without loss of generality we can assume that the computational basis is the optimal one. Otherwise we can make a change of basis in (\ref{phis}). Hence the LE is given by:
 \be \label{Entav}
\L_{ij}^{E} (\psi_{\mathrm{MP}}) =\sum_{\{s\}} p_{\{s\}} E( \tilde{\phi}_{\{s\}}  ) \ , \ee
where
$p_{\{s\}}=\la \phi_{\{s\}}|\phi_{\{s\}} \ra/ \la \psi_{\mathrm{M P}}
|\psi_{\mathrm{MP}}\rangle$ is the probability for  obtaining the normalized
state $| \tilde{\phi}_{\{s\}}\ra=|\phi_{\{s\}}\ra /  \la \phi_{\{s\}}|\phi_{\{s\}}
\ra^{1/2}$. Though the MPS representation allows us to efficiently calculate the  states after the measurements we still face the problem that the  sum (\ref{Entav}) involves an exponential number of terms  ($d^{N-2}$). To find a good approximation of this sum we propose a scheme based on the Monte Carlo (MC) method, which will be now explained in more detail.
\subsection{Monte Carlo method}
 The Monte Carlo method provides an efficient way of  selecting $M$  states $|\phi_\mu \ra$ sequentially from the (given) probability
distribution $\{ p_{\{s\}} \}$. The LE  can thus be approximated
by:
 \be \label{Entavmc} L_{(MC),ij}^{E} (\psi_{\rm{MP}})\approx \frac{1}{M} \sum_{\mu=1}^{M} E(
\tilde{\phi}_{\mu}  ) \ \pm \ \frac{1}{\sqrt{M}} \  \delta L_{ij}^E (\psi_{\rm{MP}}).
 \ee
Note that the accuracy of the MC method depends on the entanglement fluctuations $\delta L_{ij}^E $, which can be computed within the MC scheme as well. \\
For selecting the states $
|\phi_\mu \ra$  we follow the Metropolis algorithm
\cite{Metro53} and use single-spin-flip dynamics. We start with an initial state $
|\phi_\mu \ra$ corresponding to a specific measurement outcome $\{s\}$. >From this
we create a trial state  $ |\phi_\nu \ra$ by randomly picking a site $i$ and
changing the state of this spin with equal probability according to $s_i
\rightarrow \rm{mod} (s_i\pm 1, d)$ for $s_i=0,1, \ldots d-1$. For a spin-1/2
this simply amounts to a spin flip. The probability $p_\nu$ for obtaining the
trial state after a measurement can conveniently be computed using the MPS
representations (\ref{phis}). The trial state is accepted with probability:
 \be
P(\mu \rightarrow \nu)= \left\{  \begin{array}{ccc} \frac{p_\nu}{p_\mu} &
&\rm{if}\  p_\nu < p_\mu \ ,  \\  1 &  &\rm{else } \  .
 \end{array} \right.
\ee
If the trial state is accepted it serves as a starting point for creating a
new trial state. After $N-2$ steps, defining one MC sweep, the entanglement of the
current state is calculated \cite{comMC}. After $M$ sweeps the algorithm stops and the average
(\ref{Entavmc}) is performed.
\subsection{Finding the optimal measurement basis}
The definition of the LE (\ref{LE})
requires an optimization over all possible  measurement strategies.
A good guess for the optimal basis can typically by found using exact diagonalization for small system sizes, followed by numerical maximization of the average entanglement.
Alternatively, the optimal basis can also be
extracted directly from the MPS matrices  $A_i$ in (\ref{phis}). Using a generalization of the
concurrence for pure bipartite $D \times D$ states, it has been shown in \cite{VMC04}
(for an open chain with $D$-dimensional spins at the ends), that the optimal basis
is the same basis, that maximizes the expression $\sum _{s_i}
|\det(A_i^{s_i})|^{2/D}$. Hence we can consider the matrices $A_i$ as
(unnormalized) pure $D\times D \times d$ states, for which we want to calculate the LE
with respect to the $D \times D$ system. Since we measure only on a single site
this problem is equivalent to calculating the EoA of the reduced ($D^2 \times D^2$) density matrix, which can be done numerically (see also Appendix C).  For the models we studied our numerical analysis indicates that
these findings hold independent of the choice of both the entanglement measure  and the boundary conditions.
As a further numerical result we find that the optimal basis appears to be independent of the system size. Hence exact diagonalization and numerical optimization for small systems usually provides the most efficient way to find the optimal measurement strategy.
\subsection{Determination of the MPS}
In the following we are interested in   finding a good approximation of the
true ground state in terms of MPS (\ref{MPS}) for systems with PBC. We choose PBC  in order to minimize boundary effects and to better mimic the behavior in the thermodynamic limit already for small system sizes.
Let us now introduce a method to determine  translationally invariant MPS. It is based on the following idea: We use the  DMRG algorithm  for an infinite chain to extract a site-independent set of matrices $A^s$, defining a translationally invariant MPS (\ref{MPS}) for infinite $N$. We then us the same set   $A^s$ to construct a MPS with PBC for arbitrary $N$. It is obvious that this method is very efficient, particularly for large $N$, because we have to run the DMRG only once to obtain the MPS representation for any system size. One might expect that this increase in efficiency happens at the  cost of precision.  However,  an optimum in accuracy on the part of the MPS is not crucial for the numerical calculation of the LE, since the limiting factor for the accuracy is typically the MC method.

We start by briefly  reviewing
  the variant of DMRG,  represented by
$B \bullet B$ \cite{D97}, for an infinite
1D chain.
 At some
particular step the chain is split into two blocks and one spin in between. The
left block ($L$) contains spins $1,\ldots,M-1$, and the right one ($R$) spins
$M+1,\ldots,N$. Then a set of $D\times D$ matrices $\tilde{A}^{s_M}$ is determined
such that the state
 \be
 \label{Amatrix}
 |\Psi\rangle = \sum_{s=1}^d \sum_{\alpha,\beta=1}^D
 \tilde{A}^{s_M}_{\alpha,\beta} |\alpha\rangle_L \otimes |s\rangle_M
 \otimes |\beta\rangle_R,
 \ee
minimizes the energy. The states $|\alpha\rangle_{L,R}$ are orthonormal, and have
been obtained in previous steps. They can be constructed using the recurrence
relations
 \be
 \label{recurrence}
 |\alpha\rangle_L = \sum_{\alpha'=1}^D \sum_{s=1}^d
 U^{[M-1],s}_{\alpha,\alpha'} |s\rangle_{M-1}\otimes
 |\alpha'\rangle_{L'},
 \ee
where the block $L'$ contains the spins $1,\ldots,M-2$.

Numerically we find that the matrices $U^{[k],s}$ can be chosen in such a way
that they converge to  (site independent) matrices $U^s$ at the fix point of the
DMRG algorithm. Applying an appropriate transformation $R$,  these matrices  $U^s$
can be used to construct a translationally invariant MPS (\ref{MPS}) with $A^s=R
U^s R^{-1}=R  \tilde{A}^s R^T$ \cite{comMPS}. Note that starting from these states, it is
possible to calculate expectation values of products of local observables
\cite{RomerPRL, D97, Fannes}, since
 \be
 \label{Exp_Val1}
 \langle \Psi|O_1\ldots O_N|\Psi\rangle = {\rm Tr}
 \left(E^{[1]}_{O_1} \ldots E^{[N]}_{O_N}
 \right),
 \ee
where
 \be
 \label{Exp_Val2}
 E^{[k]}_{O}= \sum_{s,s'=1}^{d} \langle s|O_k|s'\rangle
 A^{s}_k\otimes \left(A_k^{s'}\right)^\ast.
 \ee
We applied this method to compute  ground state expectation values using expression (\ref{Exp_Val1}) for  various 1D spin systems with PBC \cite{comMPS1}. Our numerical analysis of  systems with finite $N$  shows that  both the energy and the  correlations  can be computed rather accurately.  The achieved accuracy is several orders of magnitude higher than  finite size effects, but also several orders of magnitude lower compared to the variational method \cite{DMRGperiodic}, which has been introduced recently. These findings even  hold for system sizes as low a $N\approx 10 $, and are rather surprising, because the MPS is constructed from the infinite chain. We further checked that also the long range behavior of correlations and entanglement is reproduced correctly by our translationally invariant MPS.

To sum up,   our numerical results indicate  that translationally invariant states can  be sufficiently well  approximated by a single set of MPS  matrices $A^s$ for almost arbitrary system size $N$. This observation, together with the findings in \cite{VMC04}, might shed some light on our previous  numerical finding that the optimal measurement basis for LE is typically both site-and size-independent.

\section{Examples of LE in spin-$1/2$ models}
In this section we apply the concept of LE  to  quantify the localizable ground state
entanglement of various spin-1/2 models. After some general considerations we compute the LE as measured by the concurrence $L_{i,j}^C$  numerically for two specific examples.
\subsection{General considerations}
We consider spin-1/2 Hamiltonians of the form
\begin{equation} \label{Hs1/2}
H=- \sum_{i,j} \ \sum_{\alpha=x,y,z} \gamma_{\alpha}^{ij}
\sigma_\alpha^i \sigma_\alpha^j - \sum_i \gamma^i \sigma_z ,
\end{equation}
 with parity symmetry, $[H,\Pi_z]=0$ and
$\Pi_z:=\otimes_{i=1}^N\sigma_z^i$. Extensive numerical calculations on systems of
up to 20 qubits showed that our lower bound is always close to the LE as measured by concurrence $L_{i,j}^C$, and
typically is exactly equal to it: this is surprising and
highlights the thightness of the given lower bound. Note also that whenever  parity
symmetry is present, the upper and lower bound are given as follows \cite{VPC04}:
\begin{eqnarray}
\hspace*{-3mm} \max\left(|Q_{xx}^{ij}|,|Q_{yy}^{ij}|,|Q_{zz}^{ij}|\right)&\leq&
L_{i,j}^C\leq\frac{\sqrt{s_+^{ij}}+\sqrt{s_-^{ij}}}{2},\\
s_{\pm}^{ij}= \left( 1 \pm \la \sigma_z^i\sigma_z^j \ra \right)^2&-& \left(\la
\sigma_z^i \ra \pm \la \sigma_z^j \ra\right)^2 .\nonumber
\end{eqnarray}
The fact, that the lower bound is usually tight, can also be derived from the
numerical observation that for Hamiltonians of the form (\ref{Hs1/2}) measurements
in the (standard) $\sigma_z$-basis ($\mathcal{M}=\mathcal{Z}$) yield in most cases the optimal result.
Expanding the ground state in that basis, $|\psi\ra_0=\sum c_{i_1 \ldots i_N} |i_1
\ldots i_N \ra$, it is straightforward to show that e.g. in the case of constant
(site independent) coupling $\gamma_x$ and $\gamma_y$ the ground state energy is
minimized if all expansion coefficients $c_{i_1 \ldots i_N}$ have the same sign.
This guarantees, together with the parity symmetry, that the average entanglement
$L_{i,j}^{\mathcal{Z},C} $ for measurements in the standard basis is equal to either the $ x-x$
or $y-y$ correlation. To be more precise we distinguish the following cases:
\begin{subequations} \label{LzQ}
\begin{align}
(\gamma_x^{ij}-\gamma_y^{ij})( \gamma_x^{ij}+ \gamma_y^{ij}) &\geq 0: \quad
L_{i,j}^{\mathcal{Z},C}=   |\la \sigma_x^i \sigma_x^j \ra|, \label{LzQxx}\\
(\gamma_x^{ij}-\gamma_y^{ij}) (\gamma_x^{ij}+ \gamma_y^{ij}) &\leq 0 :\quad
L_{i,j}^{\mathcal{Z},C}=   |\la \sigma_y^i \sigma_y^j \ra|, \label{LzQyy}
\end{align}
\end{subequations}
where the conditions refer to all sites $i$ and $j$ of the chain.
Most  of the prominent spin Hamiltonians studied in
literature, like the Heisenberg, XY or XXZ model etc., trivially fulfill one of the
conditions (\ref{LzQ}), because their coupling coefficients are site-independent. Hence,  measurements in the standard basis would yield localizable quantum
correlations that are completely determined by  classical correlations.

\subsection{Ising model}
\begin{figure}[h]
\begin{center}
      \epsfig{file=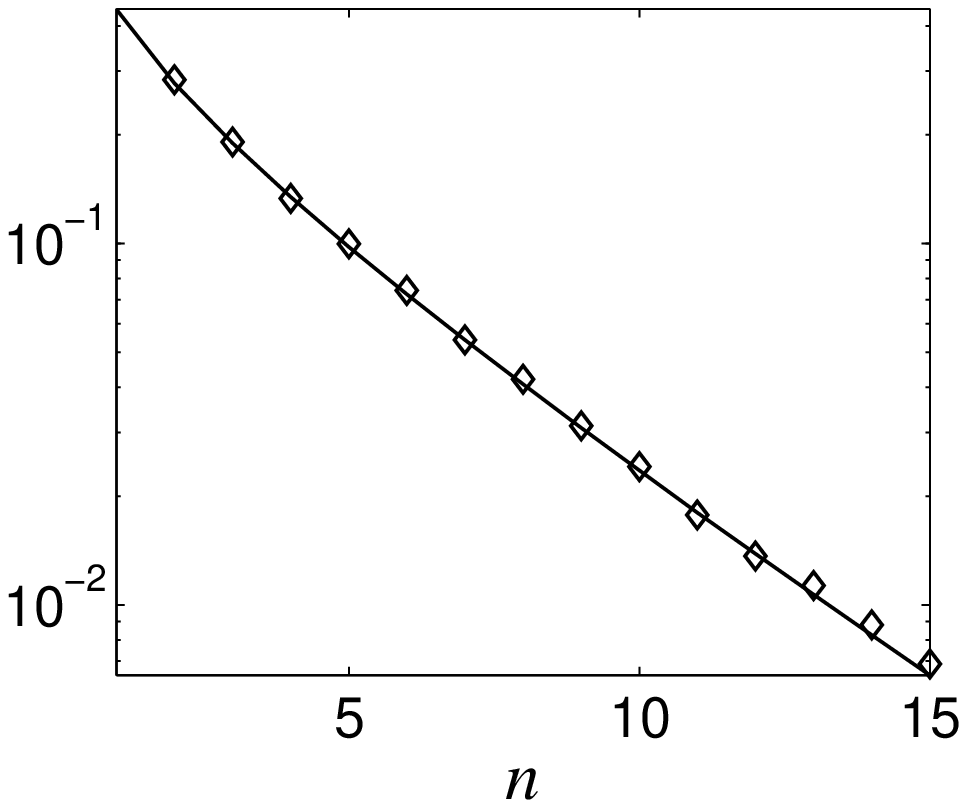,angle=-90,width=0.49\linewidth}
      \epsfig{file=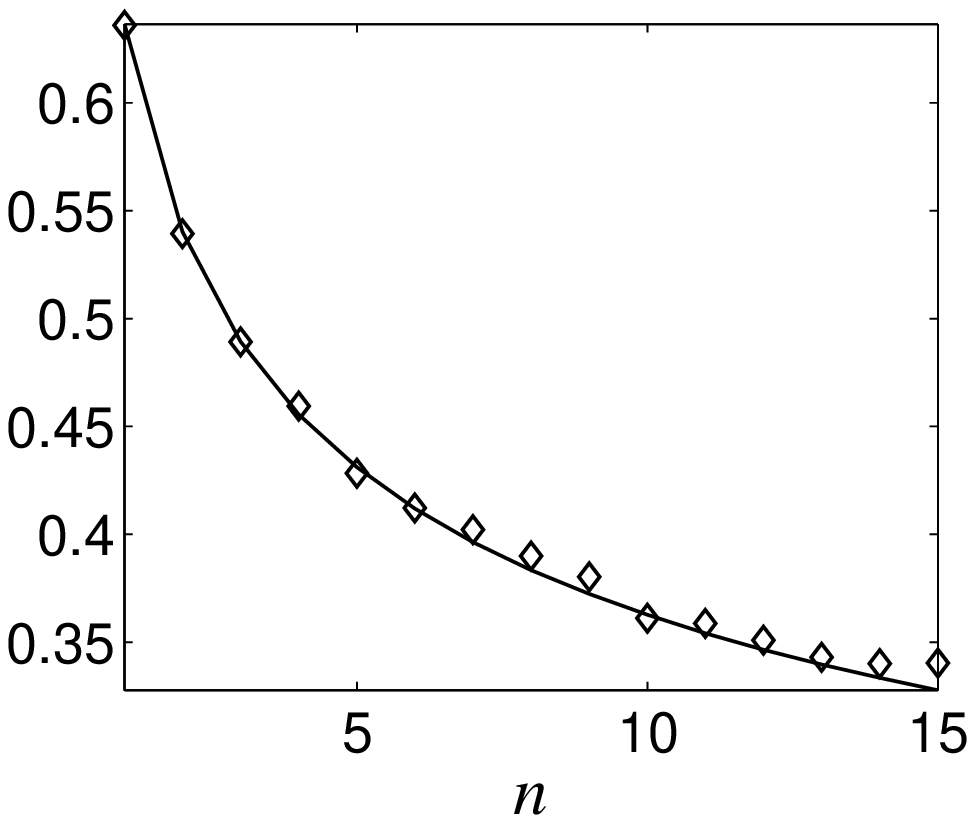,angle=-90,width=0.49\linewidth}
\end{center}
\caption{Calculation of  the  $L^C_{i,i+n} $(diamonds) for the ground state of the
Ising chain in a transverse field \mbox{(
$\gamma_{\alpha}^{ij}=\lambda\delta_{\alpha,x}\delta_{j,i+1};\gamma^i=1$ }in
(\ref{Hs1/2})) as a function of the spin distance $n$. For comparison we plot the
exact result \cite{P70} for the correlation function $Q_{xx}^{i, i+n}$. Left:
$\lambda =0.8$, exponential decrease; Right: critical point $\lambda =1$,
power law decrease ($ \sim n^{-1/4}$);  Numerical parameters (see Sect. IV):
$N=80$, $D=16$, MC sweeps $M=20,000$.}
       \label{fig-LEising}
\end{figure}

As an illustration, let us now discuss the LE of  the Ising model in a transverse
magnetic field
\mbox{($\gamma_{\alpha}^{ij}=\lambda\delta_{\alpha,x}\delta_{j,i+1};\gamma^i=1$}
in (\ref{Hs1/2})), which has been solved exactly \cite{P70} and exhibits a quantum
phase transition at $\lambda=1$. In this case, the maximal connected correlation
function is always given by $Q_{xx}$, which thus yields the best lower  bound on
LE. Numerical optimization for a finite chain indicates that the standard  basis
is
indeed the optimal one and thus the lower bound is equal to  $L_{i,j}^C$ \cite{VPC04}. We checked analytically, using perturbation theory, that for an infinite chain this numerical result is indeed true. However, for a spin distance  $n=|i-j|$ one has to go to $ n$-th order  perturbation theory, limiting this analytical treatment to rather small $n$. \\
Consequently, we can use exact results for the connected correlation function
$Q_{xx}$ \cite{P70}  to completely characterize the behavior of the LE in the
Ising chain. The Ising system is therefore also an ideal candidate for testing the
performance of our numerical method outlined in \mbox{Sect. IV.} In Fig.
\ref{fig-LEising} we plot $L^C_{i,i+n} $ and  $Q_{xx}^{i, i+n}$ as a function of
the spin distance $n$ for a chain with $N=80$ sites.

\begin{figure}[h]
        \begin{center}
      \epsfig{file=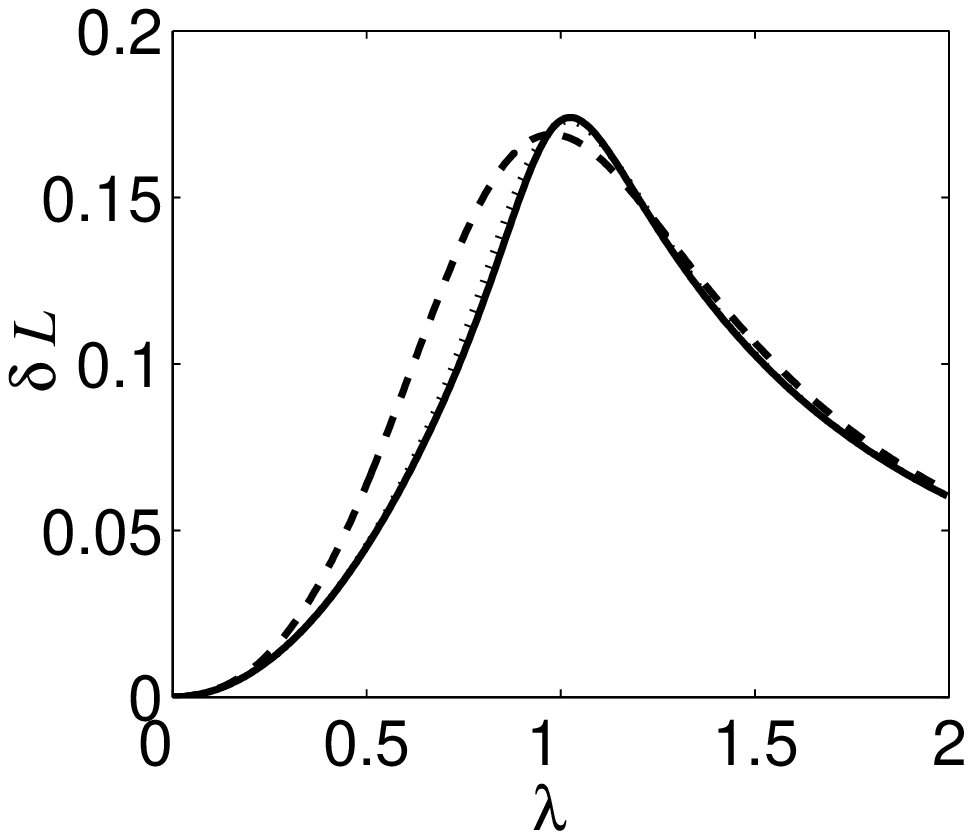,angle=0,width=0.49\linewidth}
      \epsfig{file=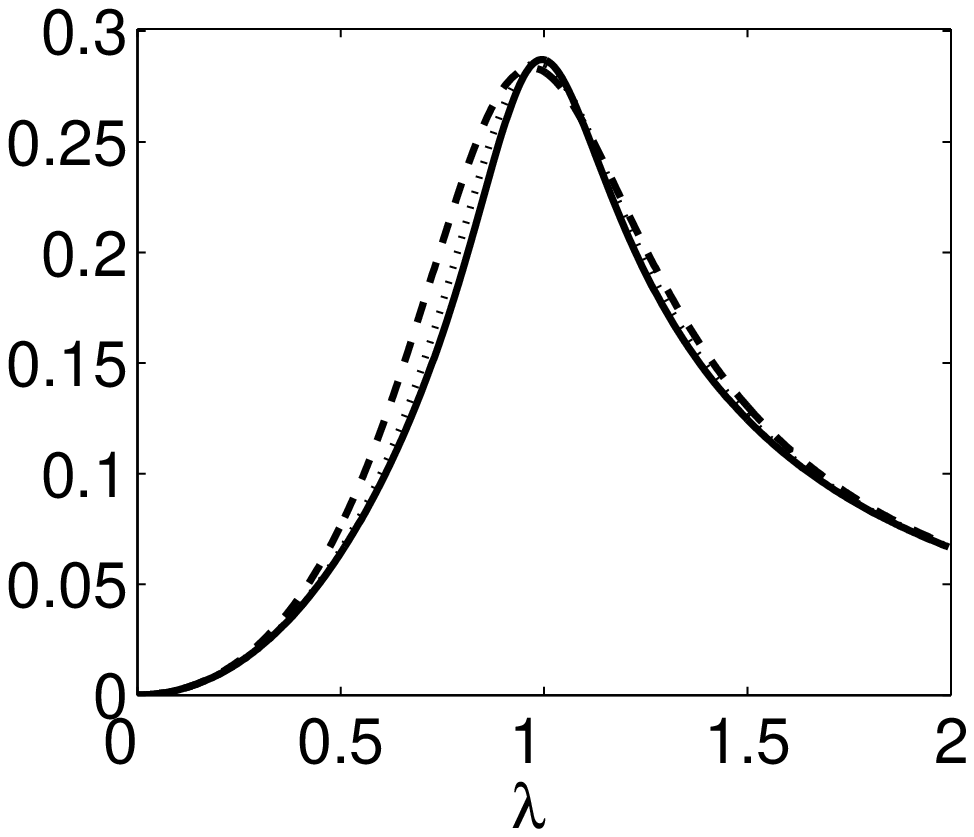,angle=0,width=0.49\linewidth}
       \end{center}
\caption{Exact calculation of entanglement fluctuations  $\delta L^C_{i,i+n}$
 (\ref{def-fluc}) as a
function of the coupling parameter $\lambda$ for a finite Ising chain in a
transverse magnetic field with PBC. Left: distance $n=1$, $N=6$ (dashed),  $N=12$ (dotted), $N=16$ (solid); Right: distance $n=4$, $N=9$ (dashed),  $N=12$ (dotted), $N=16$ (solid).}
       \label{fig-fluc}
\end{figure}

For $\lambda<1$, the LE decreases exponentially with $n$, and the entanglement length is finite. At the quantum critical point
$\lambda=1$, the behavior of the LE changes drastically, because it suddenly  decreases as a power law, $L^C_{i,i+n} \sim n^{-1/4}$, thus leading to a diverging entanglement
length $\xi_E$. In Fig. \ref{fig-LEising} we observe that  the MC
method becomes less accurate at the critical point. As we will see later, one reason is that the statistical error due to entanglement fluctuations becomes rather large at the critical point (see Fig. \ref{fig-fluc}). Another  (systematic) error might be induced by the
single-spin-flip dynamics used to create the trial state. Better results for the
critical region could possibly be achieved by applying the Wolff algorithm
\cite{W89}. Here, a cluster of spins depending on their spin orientation is
flipped, which accounts for the formation of domains.

In \cite{VPC04}  it was shown that for  the case $\lambda > 1$ we also get
$\xi_E=\infty$, since the LE saturates to a finite value given by
$M_x^2=1/4(1-\lambda^{-2})^{1/4}$. Indeed, the ground state is then close to the
GHZ-state. In a more realistic setup, however, the parity symmetry of the Ising
Hamiltonian will be broken by a perturbation and the ground state for large
coupling will also be separable, as it is given by a superposition of two GHZ
states with different parity \cite{Murg}.

Let us now study the behavior of entanglement  fluctuations $\delta L^C_{i,j}$ inherent to the
 statistical definition
of LE (see Def. (\ref{def-fluc})) as a
function of the coupling $\lambda$. In Fig. \ref{fig-fluc} we plot  $\delta L_{i,j}^C$ for different parameters  $n$ and $N$.

The maximum of the fluctuations is always located in the vicinity of the critical point $\lambda=1$ and gets shifted to larger $\lambda$ values with increasing $N$. Thus the increasing entanglement fluctuations reflect very well the increasing complexity of the wavefunction close to the critical region. The location of the maximum $\lambda_m$ in the thermodynamic  limit ($N\rightarrow  \infty$) apparently depends on the distance $n$ of the two spins. For nearest neighbors ($n=1$) we observe that the maximum of  $\delta L^C_{i,i+1}$ is somewhat shifted  to the right of the critical point ($\lambda_m\approx1.025$). For all distances $n>1$, however, our numerical calculations show that the maximum is positioned at $\lambda <1$ but becomes asymptotically close to the critical point with increasing $n$ (for $n=4$ see Fig. \ref{fig-fluc}).  Furthermore, in Fig.  \ref{fig-fluc} we see that the absolute value of the maximum increases with $n$ and becomes comparable with $L_{ij}^C$ itself \cite{comfluc}. The strong fluctuations inherent to the Ising model lead to large statistical errors in the  numerical calculation of LE using Monte Carlo (see (\ref{Entavmc})). The errors become even more pronounced for the calculation of the fluctuations. This is the main reason, why we have restricted ourselves here to exact calculations for a small system with PBC. However, we confirmed  that the data for $N=16$ represents the behavior in the large $N$ limit reasonably well and no qualitative changes occur.

\subsection{XXZ model}
\begin{figure}[h]
        \begin{center}
      \epsfig{file=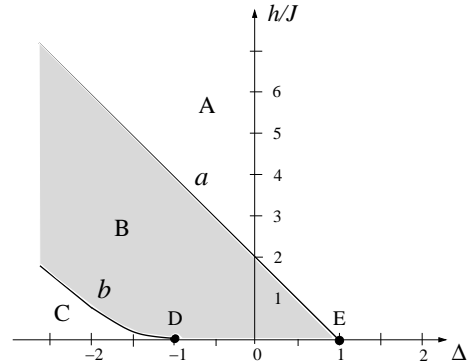,angle=0,width=0.7\linewidth}
       \end{center}
\caption{Schematic drawing of the phase diagram of the XXZ model (Eq.
(\ref{HXXZ})) as a function of the anisotropy $\Delta$ and the magnetic field $h/J$
 \cite{Takahashi}. In regions A and C the ground state has an energy gap, whereas
in region B the system becomes gapless (critical). Point E is the ferromagnetic XXX point and
point D corresponds to the antiferromagnetic XXX point. }
       \label{phasediagram}
\end{figure}

Let us now turn to the discussion of another exactly solvable 1D spin system, the
so called XXZ model \cite{Takahashi}. This model not only appears in condensed
matter physics in the context of ferro- or antiferromagnetic materials. Recently
it has been shown that it can also effectively describe the physics of ultra cold
atoms in a deep optical lattice \cite{OLspin}. The Hamiltonian can be written as
{\small
\be \label{HXXZ} H_{\rm XXZ}= - \sum_i \ [ \sigma_x^i \ \sigma_x^{i+1} + \sigma_y^i \
\sigma_y^{i+1} + \Delta \ \sigma_z^i \ \sigma_z^{i+1} + \frac{h}{J}   \sigma_z^i
 \ ],
\ee}
\hspace*{-2mm} where we have introduced two dimensionless parameters, which can be varied
independently: the anisotropy $\Delta$ and the magnetic field $h/J$ in units of
the exchange coupling. The phase diagram of the XXZ model as a function of these
two parameters \cite{Takahashi} is depicted in Fig. \ref{phasediagram}.

 The XXZ model
can be solved exactly using the Bethe ansatz \cite{Bethe}. Unfortunately analytical expressions
for the correlation functions, which would yield lower bounds for LE, have only
been worked out in special cases. E.g. for the antiferromagnetic XXX model in a
magnetic field (line of constant $\Delta=-1$ in Fig. \ref{phasediagram}) analytical solutions for the correlations are summarized in \cite{JK03}.
Numerics on a finite chain of up to 14 spins show that again measurements in the
$\sigma_z$-basis appear to be optimal. Using the result (\ref{LzQxx}) this implies that $L_{i,i+n}^C=\la \sigma_x^i \sigma_x^{i+n} \ra
$. Hence the bounds given in \cite{JK03} are tight. In particular this means that the entanglement length is zero for $h/J> 4$, and  infinite for $h/J\leq 4$.

After this introductory remarks let us study the ground state entanglement properties of the XXZ model in more detail. In particular, we are interested in finding characteristic features in the LE at the quantum phase transitions indicated by the lines $a$ and $b$ in the phasediagram Fig. \ref{phasediagram}. For this purpose we  calculate in the following the LE (for fixed $n$) numerically, using exact diagonalization, as a function of the two parameters $\Delta$ and $h/J$.
\subsubsection{LE as a function of the magnetic field $h/J$}
\begin{figure}[h]
%
      \epsfig{file=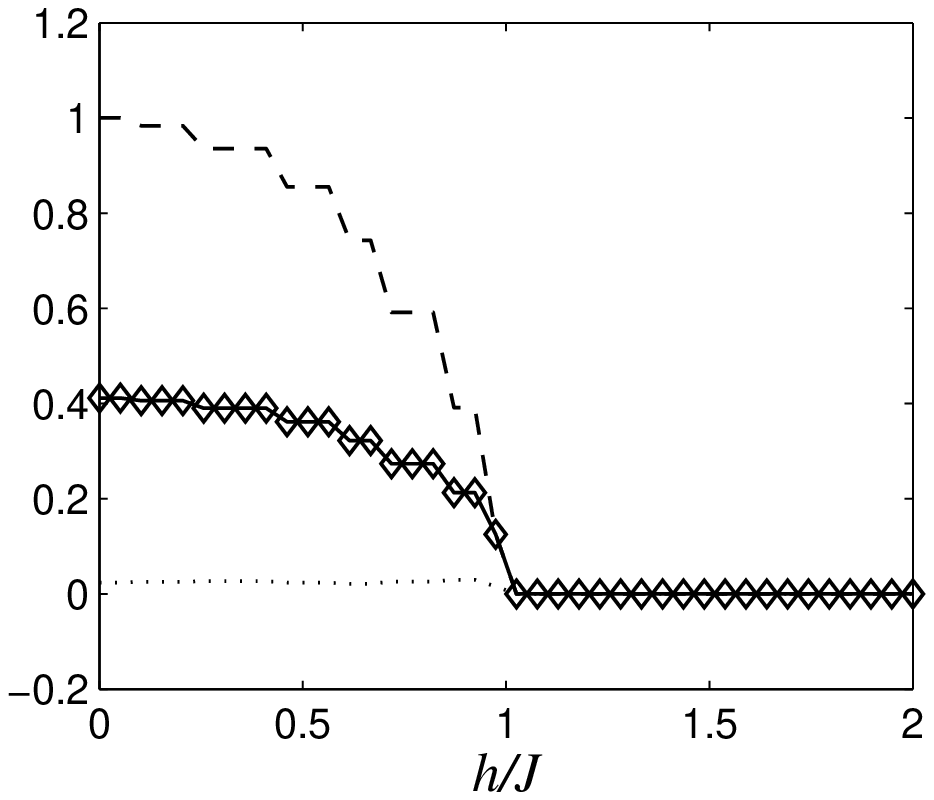,angle=0,width=0.6\linewidth}
       \epsfig{file=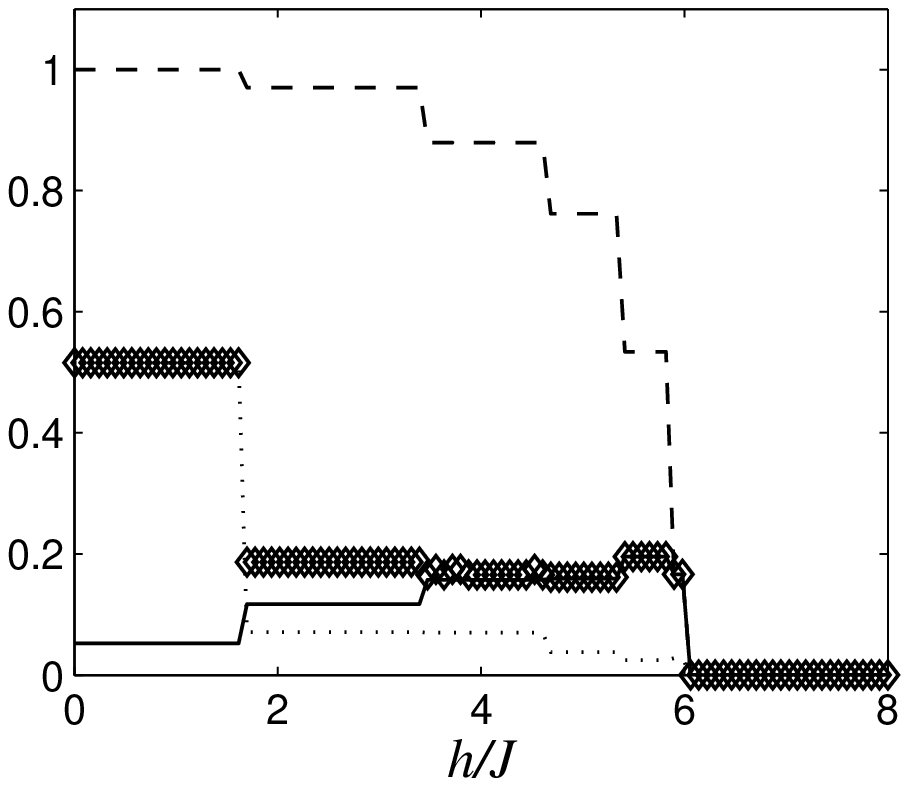,angle=-90,width=0.6\linewidth}
\caption{Calculation of  the localizable entanglement   $L_{i,i+n}^C$ (diamonds), the upper bound given by the EoA
(\ref{EoA}) (dashed) and the lower bounds $Q_{xx}^{i i+n}$ (solid) and  $Q_{zz}^{i
i+n}$ (dotted) as a function of the field $h/J$ for the  ground state of the XXZ model (\ref{HXXZ}).
The numerical calculation is performed using exact diagonalization of a chain  with  $ N=16$ sites and PBC. The distance of the two spins is  $ n=4$ sites; Up: $\Delta=0.5$; Down: $\Delta=-2$; }
       \label{fig-XXZ}
\end{figure}

In Fig. \ref{fig-XXZ} we plot  $L^C_{i,i+1}$ and the lower bounds $Q_{xx}^{ij}$ and  $Q_{zz}^{ij}$ as a function of the field $h/J$ for fixed  $\Delta=0.5$. We find that  $L_{i,i+n}^C=\la \sigma_x^i \sigma_x^{i+n} \ra$. This result can be understood as a consequence of the numerical observation, that the  standard basis appears to be optimal in the entire half space $\Delta \geq -1$.
At the critical point the LE becomes zero, because the ground state is given by a product state. Note that this phase transition (indicated by line $a$ in Fig. \ref{phasediagram}) is sharp even for finite systems, since it is due to  level crossing \cite{Takahashi}.

We now investigate
the region $\Delta<-1$, which contains a second quantum phase transition  (indicated by line $b$ in Fig. \ref{phasediagram}). The dependence of the LE and its bounds on the magnetic field for fixed $\Delta=-2$ is depicted in  Fig. \ref{fig-XXZ}.  We again observe a sharp phase transition at $h/J \approx 6$, when the system enters the unentangled phase A (Fig. \ref{phasediagram}). At $h/J \approx 1.6$  the LE  experiences a sudden drop-off. However, it remains to be checked, whether this step is due to finite size effects or a  characteristic feature of a quantum phase transition.  Note that for $h/J \lesssim 1.6$ we obtain that  $L_{i,j}^C=Q_{zz}^{ij}$. This feature can be understood by considering the limiting case of zero field and \mbox{$\Delta \rightarrow -\infty$}. There, the  Hamiltonian $H_{\mathrm{XXZ}}$ commutes with the
parity operator in $x$-direction, \mbox{$\Pi_x=\bigotimes_{i=1}^N \sigma_x^i$}, and the (doubly degenerate) ground
state for even $N$ is given by \mbox{$|\psi \ra= 1/\sqrt{2} \ (|0101...01\ra \pm
|1010...10\ra)$}. After suitable projective measurements in the \mbox{$x$-direction} these
states reduce to maximally entangled Bell states \mbox{$|\Psi^\pm \ra=  1/\sqrt{2}\
(|01\ra \pm |10\ra)$}. Hence in this limit\mbox{ $L_{i,j}^C=Q_{zz}^{ij}\rightarrow 1$} and
the $\sigma_x$-basis turns out to be the
optimal one.
  This line of reasoning strictly holds only  in the limit $\Delta \rightarrow -\infty$, but can
qualitatively be extended to the whole region C (Fig. \ref{phasediagram}).
\\
For  $h/J \gtrsim 1.6$ we observe in Fig. \ref{fig-XXZ} one of the rare examples, for which the LE is not exactly equal to the maximum correlation function. Note also that in this region the maximum correlation function changes from $Q_{zz}^{ij}$ to  $Q_{xx}^{ij}$. When approaching the critical point $h/J \approx  6$ the standard basis becomes close to optimal again and  $L_{i,j}^C\approx Q_{xx}^{ij}$. \\
In the following we have a closer look on the phase transition between the regions B and C in Fig. \ref{phasediagram}.

\subsubsection{LE as a function of the anisotropy $\Delta$}
\begin{figure}[h]
      \epsfig{file=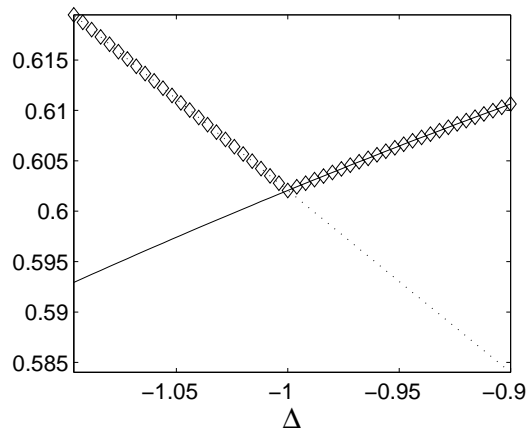,angle=0,width=0.8\linewidth}
\caption{Calculation of  $L^C_{i,i+1}$ (diamonds) and the lower bounds $Q_{xx}^{i i+1}$ (solid) and  $Q_{zz}^{i
i+1}$ (dotted) as a function of the anisotropy $\Delta$ for the  XXZ model (\ref{HXXZ})  with zero field $h/J$. At the  critical point $\Delta=-1$ the  LE exhibits a cusp, in contrast to the monotonic behavior of the  correlation functions.
The calculation is performed using exact diagonalization and numerical optimization for a chain  of length  $ N=10$  with periodic boundary
conditions. }
       \label{fig-XXX}
\end{figure}

We consider the case of zero magnetic field and study the LE as a function of $\Delta$ in the vicinity of the antiferromagnetic XXX point. At the critical point $\Delta=-1$ the ground state undergoes a Kosterlitz-Thouless \cite{Sachdev} quantum phase transition. In Fig. \ref{fig-XXX} we have calculated numerically the localizable entanglement $L^C_{i,i+1}$ for nearest neighbors  and the corresponding lower bounds $Q_{xx}^{i,i+1} $ and  $Q_{zz}^{i,i+1}$.

One sees that  $L^C_{i,i+1}$ is equal to the maximum correlation function. However, at the critical point the maximum correlation function changes, due to the crossing of   $Q_{xx}^{i,i+1} $ and  $Q_{zz}^{i,i+1}$, thus  leading to a cusp in the LE. Hence the quantum phase transition is characterized by a  discontinuity in the first derivative of $L^C_{i,i+1}$. This result is remarkable, because  for this Kosterlitz-Thouless transtion the ground state energy (i.e nearest neighbor correlation functions) and all of its derivatives are continous \cite{KT-XXX}. As shown in \cite{XXX-C1,XXX-C2},  the concurrence (or the derivative) of the reduced density matrix $\rho_{i,i+1}$  also does not exhibit a discontinuity at the critical point. This result is expected, because the concurrence is a function of one-particle and two-particle correlation functions.  We further note that, according to our numerical analysis the reason for the cusp in  $L^C_{i,i+1}$ is that the optimal measurement basis changes at the critical point abruptly from the $\sigma_z$- to the $\sigma_x$-basis.

So far this is a  purely numerical finding for a finite dimensional system. However, given that the average entanglement can be maximized by applying the same unitary transformation on  all spins, one can rigorously show that indeed a  cusp in the LE  must occur exactly at the critical point and independently of the size $N$. The argument goes at follows: At the antiferromagnetic XXX point the Hamiltonian (\ref{HXXZ}) possesses $SU(2)$ symmetry. This means that any measurement basis yields the  same LE. In particular, we know from (\ref{LzQxx}) that for measurements in the standard basis the LE is equal to the  correlation function $Q_{xx}^{i,i+1}$. At the critical point we thus have:  $L^C_{i,i+1}=Q_{zz}^{i,i+1}=Q_{xx}^{i,i+1}$.
 Since connected correlation functions yield a lower bound to  $L^C_{i,i+1}$ (see Sect. III) the localizable entanglement  $L^C_{i,i+1}$ must exhibit a cusp at the critical point, where  $Q_{zz}^{i,i+1}$ and $Q_{xx}^{i,i+1}$ cross.

In summary our discussion of the LE in various spin-1/2 models has shown in which
parameter regimes these systems can be used for e.g. localizing long-range
entanglement as indicated by the entanglement length.
 We have further seen that the study of   LE as well as the entanglement fluctuation  provides a   valuable tool for  detecting and characterizing quantum phase transitions. In addition our numerical results  indicate that the ground state
entanglement of spin-1/2 Hamiltonians with two-spin nearest neighbor interactions
is typically very well
described by the maximum correlation function. We will now see that this
observation is not necessarily true for spin-1 systems.

\section{LE  in the Spin-1 antiferromagnetic Heisenberg chain}
We study the ground state entanglement of  the  (generalized) antiferromagnetic   Heisenberg
chain:
\be \label{HeisAF}
H_{AF} =  \sum_{i=1}^{N-1}\left[
  \vec{S}_{i}\cdot \vec{S}_{i+1} -
  \beta (\vec{S}_{i}\cdot \vec{S}_{i+1})^2 \right] ,
 \ee
which includes a biquadratic term.
In a recent work \cite{JuanjoAKLT} it has been demonstrated that quantum Hamiltonians of this kind can be implemented with ultra cold atoms trapped in an optical lattice potential.
\subsection{General considerations}
For the  Heisenberg antiferromagnet (AF) ($\beta=0$) it has been shown by Haldane \cite{H83} that in the case of half-integer spins
the spectrum is gapless in the thermodynamic limit,  and thus the correlation length of the ground state is
infinite.    For integer spins, however,  an energy gap emerges, resulting in a
finite correlation length.  Let us  now investigate whether a similiar connection holds for
the LE and the corresponding entanglement length.\\
>From the  lower bound (\ref{LElb}) it follows that the  predicted infinite
correlation length in
spin-1/2 systems (with integer $\log_2(2 S+1)$) automatically implies a
diverging
entanglement length. However, in the case of integer spins the correlation length is finite. Hence the lower bound (\ref{LEElb}) for spin-1 systems includes the intriguing  possibility that  correlations and  entanglement may exhibit a completely different behavior. Indeed, for gapped spin-1 systems an example for this has already been found \cite{VMC04}. At the AKLT \cite{AKLT} point ($\beta=-1/3$) the entanglement length of the ground state diverges, although the correlations decrease exponentially. Since the AKLT model is closely related to the Heisenberg AF, one might expect that both systems show qualitatively the same behavior.
\subsection{Numerical study}
Let us now have a closer look on the spin-1 Heisenberg AF. Since in this case we cannot resort to an analytical solution as for the AKLT, we have to rely on numerical methods. \\
We start our analysis by performing exact diagonalizations for an open chain of
up to 10 sites. At the endpoints we couple to $ S=1/2$ spins thus making sure that the system is in the singlet ground state. We
are interested in the LE between the endpoints of the chain. Since the end spins
are represented by qubits we can still refer to the LE as measured by the concurrence: $L_{1,N}^C$. Our numerical analysis
shows that the  optimal measurement basis  is given by  the same local unitary
transformation: \be \label{UHeis} U=\frac{1}{\sqrt{2}} \left[ \begin{array}{ccc}1 & 0 & 1
\\ 0 & \sqrt{2} & 0\\ -1 & 0& 1 \end{array} \right]\ ,
 \ee
as for the AKLT.
This strategy produces a maximally entangled state between the end spins
($L_{1,N}^C=1$).  This surprising result can
be understood from the analytical study of the AKLT model
in \cite{VMC04}.  In a singlet
valence bond picture measurements in the basis (\ref{UHeis})  can be interpreted
as   Bell measurements on a  (virtual) \mbox{spin-1/2} system, which  lead to
entanglement swapping.

Let us now investigate, whether this effect depends on  the choice of the
boundary condition or the number of sites. For this purpose we apply our numerical method outlined in Sect. IV. The LE in its variant $L_{ij}^{E_E}$ \mbox{(see (\ref{EEdef}))} is computed for a chain with PBC and a large
number of sites ($N=80$).
\begin{figure}[h]
\begin{center}
      \epsfig{file=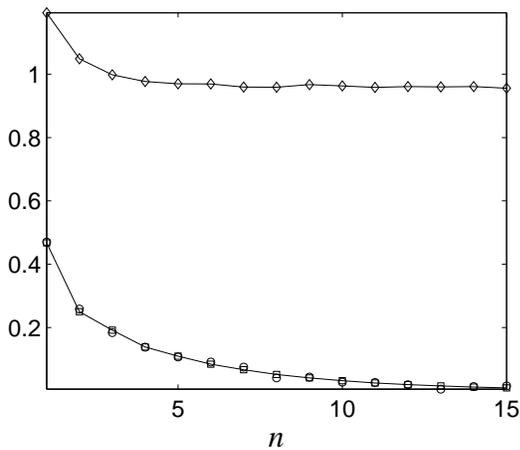,angle=-90,width=0.8\linewidth}

\end{center}
\caption{Calculation of  $
L_{i,i+n}^{E_E}$  (diamonds) for the ground state of the
antiferromagnetic spin-1 Heisenberg chain with $N=80$ sites as a function of the spin
distance $n$. For comparison we
plot the correlation function $Q_{xx}^{i, i+n}$ computed directly from the MPS
(squares) and using Monte Carlo (circles). Numerical parameters (see Sect. IV):
$D=16$, MC sweeps $M=20,000$.}
       \label{fig-LEHeismc}
\end{figure}

In Fig. \ref{fig-LEHeismc} we see that the LE saturates at a finite value $
L_{i,i+n}^{E_E}\rightarrow 0.960 \pm 0.003$ for large $n$, whereas the correlations
decrease exponentially. This demonstrates that the ground state of the
antiferromagnetic spin-1 Heisenberg chain could be used to distribute EPR-like
entanglement over arbitrary distances by performing local operations on the
intermediate spins. As mentioned in Sect. II.D this result might be particularly interesting in the context of quantum repeaters.\\
Let us now come back to Haldane's result for the  Heisenberg AF stated in the beginning of this section. Our numerical study of the spin-1 case might give a first indication that, unlike the correlation length, the entanglement length is infinite  for both half-integer and integer spins.

\subsection{Hidden order and string order parameter}
Our numerical results show that the ground state of the spin-1 Heisenberg AF exhibits long range order in terms of the localizable entanglement. We have also seen that this long range order is not reflected in the behavior of the two-particle correlation functions. However, one can define a multiparticle correlation function, the so called   \emph{string order  correlation function}
\cite{string-order}, which detects this \emph{hidden order} in the ground state.  The string order correlation function has been argued to be of topological nature and is defined as
 \begin{equation} \label{Qso}
 Q_{so}^{i,i+n}= \langle S_z^i\left[\otimes_{k=i+1}^{i+n-1} R_k \right] \otimes S_z^{i+n}\rangle,
 \end{equation}
with $ R_k=\exp\left(i\pi S_z^k\right)$. A non-vanishing \emph{string order parameter}, \mbox{$\xi_{so}:= \lim_{n\rightarrow\infty} Q_{so}^{i,i+n}$}, indicates the presence of long range (hidden) order. As an obvious generalization of the
string order correlation function to arbitrary models, let us define a connected  version  in a variational way. Consider the set of all
observables $\{\hat{O}\}$ with bounded spectrum
$-\openone\leq\hat{O}\leq\openone$. We define the \emph{connected string order
 correlation function} $Q_{cso}^{i,i+n}$ (and the related parameter $\xi_{cso}$) for a given translational invariant state as
\begin{equation} \label{Qcso}
 Q_{cso}^{i,i+n}=\max_{-\openone\leq\hat{O}_1,\hat{O}_2\leq\openone}
\langle
\hat{O}_1^i\left[\otimes_{k=i+1}^{i+n-1}\hat{O}_2^k\right]\hat{O}_1^{i+n}\rangle_{c}
.
\end{equation}
Here $\la A_1 A_2  \ldots A_n \ra_{c}$ denotes the connected
n-point correlation function, which can be defined in a recursive
way: \bea
 \la A_1 \ra & =&  \la A_1 \ra_c ,\\
 \la A_1 A_2 \ra & =& \la A_1 \ra_c \la A_2 \ra_c + \la A_1 A_2 \ra_c , \\
 \la A_1 A_2 A_3 \ra &=& \la A_1 \ra_c \la A_2 \ra_c \la A_3 \ra_c + \la A_1 \ra_c \la A_2 A_3 \ra_c \\
 &+&
\la A_2 \ra_c \la A_1 A_3 \ra_c + \la A_3 \ra_c \la A_1 A_2 \ra_c
\nonumber
\\ & + &\la A_1 A_2 A_3 \ra_c ,  \nonumber \\
& \vdots& \nonumber \eea
 Note that the
connected part assures that $ Q_{cso}^{i,i+n}$
 measures a nonlocal correlation, and that  the string order parameter of the AKLT-ground state is
indeed recovered by this definition.

It has been verified numerically that $\xi_{so}$ is finite for the
ground state of the spin-1 Heisenberg AF. This fact can rigorously
be proven for the related AKLT-ground state \cite{AKLT}. For this
state it was further  shown that the LE saturates as well with the
spin distance $n$ \cite{VMC04}. Hence one might expect a
connection between the existence of long range order in the
entanglement and long range order indicated by the string order
parameter. However, one can
find examples for which this connection does not hold.\\
 For instance, in
\cite{VMC04} it has been shown that already an infinitesimal
deformation of the AKLT model leads to an  exponentially
decreasing  LE, whereas  $\xi_{cso}$ stays finite.  On the other
hand, ground states exist that exhibit a diverging entanglement
length but vanishing $\xi_{cso}$. A simple example can be found in
the class of MPS (\ref{MPS}) defined on qubits ($d=2$) and with
qubit bonds ($D=2$). Note that all these MPS are guaranteed to be
ground states of some local Hamiltonians. Furthermore, for MPS
with qubit bonds  the string order parameter and the LE can easily
be computed analytically (see \cite{VMC04} and \mbox{Appendix C}).
In particular, let us study the state defined by \be
A^1=\sigma_z+\sigma_y\hspace{1cm} A^2=\sigma_z-i\openone.
\ee
The
entanglement length can easily be proven to be infinite. A
necessary condition for $\xi_{cso}$ to be nonzero is that there
exists a unitary operator $\hat{O}_2$ for which the largest
eigenvalue of $E_{\hat{O}_2}$ has the same magnitude as the
maximal eigenvalue of $E_{\openone}$ (Appendix C). For the example
given, this is impossible, hence providing an example of a ground
state with a diverging entanglement length but no  long range
hidden order.

\section{Mixed states}
In this section we apply the concept of LE to characterize the entanglement of
multipartite mixed states. Note that the definition of LE (\ref{LE}) already
includes the possibility of having a  mixed state $\rho$. This implies that the states  $ \rho_{\{s\}}=\la \{s\} |\rho| \{s\} \ra$ after the measurements are also mixed. Hence we  refer in the following to the LE as measured by the negativity: $L_{ij}^N$.
In order to provide a tool for the computation of $L_{ij}^N$ let us now generalize the numerical method, outlined in Sect. IV, to mixed states.

\subsection{Numerical method}
The key point is to find a
representation of a mixed state in terms of low dimensional matrices $A_k$,
analogous to the MPS (\ref{MPS}). This problem has been considered recently in
\cite{DMRGmixed}. There the concept of MPS is generalized to matrix product
density operators (MPDO), which are defined as
\begin{eqnarray}
  \label{MPDO}
  \rho = \sum_{s_1,s_1',\ldots,s_N,s_N'=1}^d {\rm Tr} (M^{s_1,s_1'}_1 \ldots
  M^{s_N,s_N'}_N) \nonumber\\
  \times |s_1,\ldots,s_N\rangle\langle s_1',\ldots,s_N'|,
\end{eqnarray}
where $M^{s_k,s_k'}_k$ are $D_k^2\times D_{k+1}^2$ matrices. They can be
decomposed as
\begin{equation}
  \label{Mpurified}
  M_k^{s,s'}=\sum_{a=1}^{d_k} A_k^{s,a} \otimes (A_k^{s',a})^\ast.
\end{equation}
The state $\rho $ can be purified into a MPS by including ancilla  states $\{ |
a_k \ra \}$ of dimension $d_k$:
\begin{equation}
 \label{MPPurification}
 |\Psi\rangle = \sum_{s_1,\ldots,s_N}\sum_{a_1,\ldots,a_N}
 {\rm Tr} \left(\prod_{k=1}^N A^{s_k,a_k}_k\right)
 |s_1a_1,\ldots,s_N a_N\rangle.
\end{equation}
In \cite{DMRGmixed} a method has been  introduced that allows one to determine  the matrices $A_k$ iteratively for a given Hamiltonian and temperature.\\
Starting from the mixed state in MPDO representation (\ref{MPDO})  the LE can be computed along the same lines, using the MC method, as for pure states. However, we note that the matrices $M_k$ in (\ref{MPDO}) have dimension $D^2\times D^2$, compared to the $D \times D$ matrices $A_k$ for pure states. Thus in the case of PBC the computation time for the MC part scales at least with $d~ D^5$ for mixed states, but only with  $ D^3$ for pure states.  \\
Let us also comment briefly on the optimal measurement basis. In contrast to pure states, the optimal basis for mixed states can no longer be deduced directly from the matrices $M_k$.
 Alternatively a good guess for the  best measurement
strategy  can be found by exact diagonalization of the Hamiltonian for small $N$, followed by numerical optimization of  LE.
\subsection{Example: AKLT model}
Next we apply this  numerical method to a specific example. Interesting candidates can be found in gapped spin-1 systems, like the  AKLT model or the Heisenberg AF (see (\ref{HeisAF})). We have seen  that the ground state of these two models  exhibits an infinite entanglement length \cite{VMC04}. Thus the natural question arises, to what extent this feature holds for small but finite temperatures and how the entanglement length scales with temperature. \\
For our numerical study  we choose the AKLT model. The reason is simply that, here, mixed states can be approximated rather well by matrices with dimensions as small as $D\approx 10$, even for very small temperatures. This makes the computation of LE much more efficient compared to the Heisenberg AF.  Efficiency is also the reason for choosing OBC.

\begin{figure}[h]
\begin{center}
%
  \epsfig{file=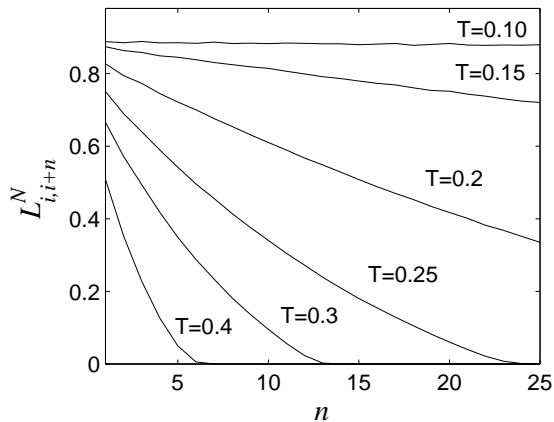,angle=-90,width=0.85\linewidth}
\end{center}
\caption{Calculation of  the LE as given by $L_{i,i+n}^N$ for the AKLT model as a  function of
the spin distance $n$  and for various temperatures $T$. We have chosen a chain with OBC and  \mbox{$N=50$} sites. We note that almost identical data
 can be obtained for \mbox{$N=20$}, indicating that our results are already close to the
thermodynamic limit. Numerical parameters (see also \mbox{Sect. IV}):
 matrix dimension $D=10$, MC sweeps \mbox{$M=5,000$}.}
       \label{fig-mixed}
\end{figure}

Let us now discuss the optimal measurement strategy. First of all we point out, that with OBC  the ground state of the AKLT is four-fold degenerate \cite{comAKLT}. Thus for $T\rightarrow 0$ the density matrix is an equal mixture of these four states, which strongly reduces the LE compared to e.g. the singlet ground state studied in \cite{VMC04}. On the other hand it is known that the degeneracy results only from the end spins of the chain. Thus, one can strongly reduce this  boundary effect by choosing the two spins, $i$ and $i+n$, to be far away from the boundaries. For this situation we found that the optimal measurement scheme for LE is given by measurements in the $U$-basis (\ref{UHeis}) on the spins between sites $i$ and $i+n$, and in the standard basis everywhere else. This result is not very surprising in terms of the valence bond picture in  \cite{VMC04}. Entanglement swapping is only needed between the two spins of interest, whereas the effect of the degeneracy can be minimized by measuring the outer spins in the standard basis. Using this strategy we  plot in Fig. \ref{fig-mixed} the LE, as given by $L_{i,i+n}^N$, depending on the spin distance $n$ for various temperatures $T$. The temperatures are chosen to be of the order of the enery gap \cite{AKLTgap}.

 The data indicates an exponential increase of  the entanglement length  $\xi_E \sim e^{\alpha/T}$ with $\alpha \approx 0.8$, thus leading smoothly to an infinite entanglement length at zero temperature. This behavior is not unexpected for a 1D system from the perspective  of the Mermin-Wagner theorem \cite{MWtheorem}.  However, it is not clear whether this theorem is really applicable to phase transitions in terms of LE. It is more inspiring to treat this problem on the basis of \emph{projected entangled-pair states} \cite{VC04}. In this picture  a finite temperature phase transition for LE  could possibly occur for two or more dimensions. We note that recently such a transition has been shown to exist for 3D cluster states \cite{R04}. \\
Finally we would like to point out that, although the entanglement length of the AKLT model is finite for $T>0$, it can still be considerably large for sufficiently low temperatures $T \lesssim 0.2$. Thus for practical purposes this system  might  still be useful, e.g. for quantum repeater setups.
\section{Conclusion}
In this article we have presented a  detailed discussion of the idea and basic properties of LE, as defined in \cite{VPC04}. Apart from that we also found new results: The central finding of  \cite{VPC04}, namely  the lower bound of LE in terms of connected correlation functions, has been generalized to pure qutrit states. Moreover we have proposed a numerical method, applicable to both pure and mixed states, that allows to calculate the LE efficiently even for large 1D spin systems. In future work we are planning to generalize this numerical  scheme  to two and higher dimensional spin systems based on the method \cite{VC04}.  We  have applied our numerical scheme to study the LE of various spin models.  The results can be summarized as follows:\\
The LE, as well as the entanglement fluctuations, exhibit characteristic features at a quantum phase transition. This result is a direct consequence of the numerical observation, that, for ground states of  spin-1/2 systems with two-spin interactions, the lower bound is typically tight, i.e. the LE is completely characterized by the maximal connected correlation function. However, connected correlation functions give only a coarse grained picture for the LE. As an example, we have shown for the spin-1 Heisenberg AF that the entanglement length diverges, whereas the correlation length is finite. This numerical result might suggest that, in terms of LE and opposed to connected correlation functions, no distinction has to be made between the scaling behavior of integer and half-integer spin Heisenberg AF's. To confirm this idea, further studies of the integer spin case are desired. \\
We further note that  preliminary results indicate that the entanglement features of the Heisenberg AF might hold qualitatively for the whole class of gapped spin-1 models defined by (\ref{HeisAF}). This finding would allow one to distinguish the Haldane phase from the dimerized phase based on the scaling of LE. Let us also mention in this context that gapped spin-1 systems also appear to be ideal candidates to look for quantum phase transitions, being detected solely by the entanglement length and not the correlation length (like for the generalized AKLT model  in \cite{VMC04}).

In order to illustrate that our numerical method works also for mixed states we computed the LE of the AKLT model for finite temperatures. We found that the entanglement increases exponentially with the inverse temperature. This smooth behavior indicates the absence of a phase transition in terms of LE for the 1D case. However, we have reason to believe that such a phase transiton might occur for the 2D system.

\section{Acknowledgements}
We acknowledge helpful discussions with  J.-J. Garc\'ia-Ripoll,
who also provided the numerical program for computing MPDO's in
the case of finite temperature. We also thank D. Loss for
suggesting to address the issue of entanglement fluctuations. This
work was supported in part by EU IST projects (RESQ and QUPRODIS),
the DFG (SFB 631), the ``Kompetenznetzwerk
Quanteninformationsverarbeitung'' der Bayerischen Staatsregierung,
the DGS under contract BFM 2003-05316-C02-01 and the Gordon and
Betty Moore Foundation (the Information Science and Technology
Initiative, Caltech).
\begin{appendix}
\section{Connected correlation functions and local measurements on pure qubit states}
Here we present an extended version of the proof in \cite{VPC04} for the following statement:\\
\emph{ Given a (pure or mixed) state $\rho$ of $N$ qubits with connected
correlation function $Q^{ij}_{A B}$ between the spins $i$ and $j$ and directions
$\vec{a}, \vec{b}$, then there always exists a basis in which one can locally
measure the other spins such that this correlation does not decrease, on average.}\\
\emph{Proof:}\\
Formally we have to show that there exists a measurement $\mathcal{M}$ such that:
\be \label{ineq1}
|Q^{ij}_{A B}(\rho^{ij})| \leq \sum_s p_s |Q^{ij}_{A B}(\rho_s^{ij})| .
\ee
To this end let us first consider  mixed states of three qubits.
A mixed 3-qubit density operator can be parameterized by four $4\times 4$ blocks
\be\rho=\left[\begin{array}{cc} \rho_1 &\sigma\\\sigma^\dagger
&\rho_2\end{array}\right].\ee

Since local unitary operations can be absorbed in $\rho$ it is sufficient to
consider the $Q_{zz}^{12}$ correlations. Thus the original correlations are
completely determined by the diagonal elements of the reduced density operator
$\rho_1+\rho_2$.
  A von Neumann measurement in the basis
\begin{eqnarray}
|+ \rangle&:=&\cos(\theta/2)|0\rangle + \sin(\theta/2)e^{
i\phi} |1\rangle , \\
 |- \rangle&:=&-\sin(\theta/2) e^{-i \phi} |0\rangle +
\cos(\theta/2)|1\rangle ,
\end{eqnarray}
on the third qubit results in the hermitian unnormalized 2-qubit operators
{\small \begin{eqnarray}
X_\pm&:=&\langle \pm |\rho| \pm \rangle=\frac{\rho_1+\rho_2}{2}\pm\cos(\theta)\frac{\rho_1-\rho_2}{2}\\
&&\hspace{0cm}\pm\sin(\theta)\left(\cos(\phi)\frac{\sigma+\sigma^\dagger}{2}
+\sin(\phi)\frac{i(\sigma-\sigma^\dagger)}{2}\right) , \nonumber
\end{eqnarray}}
with probabilities $p_\pm={\rm Tr}(X_\pm)$ conditioned on the outcome $\{+ \}$ or $\{- \}$.  From these equations we see that the
SU(2) transformation on the third qubit can be accounted for by a SO(3) rotation of
the z-axis, defined by the unit vector: \be
\vec{x}:=[\cos(\theta);\sin(\theta)\cos(\phi);\sin(\theta)\sin(\phi)] \ .\ee As
noted above we have to consider only the diagonal parts of the measurement
outcomes $X_\pm$, which can be represented in terms of the column vectors:
 \be
\vec{X}_\pm:=\frac{1}{2} \ R
\left( \begin{array}{c} 1 \\ \pm \vec{x}
\end{array}\right) \ , \ee
where $R$ is the real $4\times 4$ matrix whose columns consist of the diagonal
elements of the matrices $(\rho_1+\rho_2)$, $(\rho_1-\rho_2)$,
$(\sigma+\sigma^\dagger)$, $i(\sigma-\sigma^\dagger)$.

Provided with these definitions the inequality (\ref{ineq1}) can be written in the form:
 \[
p_+|Q_{zz}(X_+/p_+)|+p_-|Q_{zz}(X_-/p_-)|\geq|Q_{zz}(X_++X_-)|.
\]
Inserting $\one=\tr (\vec{X}_\pm/p_\pm)$, this inequality can be transformed in a
bilinear form in $\vec{x}$:
 {\small \be \label{ineq3}
 \frac{1}{p_+} \left|(1 \ \ \vec{x}^T ) S\left( \begin{array}{c} 1 \\ \vec{x}
 \end{array}\right)\right|+ \frac{1}{p_-}  \left|(1 \ -\vec{x}^T ) S\left( \begin{array}{c} 1 \\ -\vec{x}
 \end{array}\right)\right|\geq |4 \alpha| \ .
\ee}
 Here $\alpha$ is the first element of the matrix
\begin{equation}
\label{Smatrix} S:=R^T(\sigma_y\otimes\sigma_y)R=\left[\begin{array}{cc} \alpha
&\vec{\beta}^T\\
\vec{\beta} & Q\end{array}\right] \ ,
\end{equation}
and $\vec{\beta},Q$ are defined as $3\times 1$ and $3\times 3$ blocks,
respectively.
 Without loss of
generality we can assume that $\alpha$ is positive and thus remove the absolute value sign in (\ref{ineq3}). Some straightforward algebra yields then the sufficient inequality:
\begin{equation}
\vec{x}^T\left( A+B \right) x\geq
0 \ , \label{ineq2}
\end{equation}
with
\be
A:= \alpha\left(\vec{c}-\frac{\vec{\beta}}{\alpha}\right)
\left(\vec{c}-\frac{\vec{\beta}}{\alpha}\right)^T, \quad B:= Q-\frac{\vec{\beta}\vec{\beta}^T}{\alpha},
\ee
where $\vec c$ is such that $p_\pm=(1\pm \vec{c}^T \vec x)/2$. We now have to show that the matrix $A+B$ has at least one positive eigenvalue. From the form of $A$ one immediately sees that it is positive semidefinite ($\alpha > 0$). The matrix $B$ requires more work. First we note that the matrix $\sigma_y \otimes \sigma_y$ in (\ref{Smatrix}) has two negative and two positive eigenvalues.
Assuming nonsingular $R$ it follows from Sylvester's law of inertia \cite{Horn85} that
$S$ also has two positive and two negative eigenvalues \cite{com1}, and so has the inverse $S^{-1}$. Now
$B$ is the inverse of the Schur complement of
$\alpha$, and hence corresponds to a principal $3\times 3$ block of the matrix
$S^{-1}$:
\be \label{Sprin}
S^{-1}=\left( \begin{array}{cc} * & *  \\ {*} & B^{-1} \end{array} \right) ,
\ee
where the entries $*$ are of no interest here. Let us denote the eigenvalues of $S^{-1}$ in algebraic increasing order by $\lambda_1 \ldots \lambda_4$ and those of $B^{-1}$ by $\mu_1 \ldots \mu_3$. From the  interlacing properties of eigenvalues of principal blocks
\cite{Horn85}, we obtain the following relation:
\be
\lambda_1 \leq \mu_1 \leq \lambda_2 \leq \mu_2 \leq \lambda_3 \leq \mu_3 \leq \lambda_4 .
\ee
Knowing that $\lambda_3 > 0$ we deduce that $B^{-1}$ posesses at least one positive eigenvalue, and so does $B$.
The existence of one positive eigenvalue in $A+B$ ensures that one can always find a measurement direction $\vec{x}$ such that the inequality (\ref{ineq2}) is
fulfilled. We have proven the theorem for a mixed three qubit state. However, this
result can immediately be extended to arbitrary $N$. To see this let us consider
e.g. the correlation $|Q_{zz}(X_+/p_+)|$ for one of the measurement outcomes on
the third qubit. The two qubit state $X_+$ can be expanded in a basis
corresponding to a measurement of the fourth qubit ($X_+=Y_+ +Y_-$).  The theorem
can now be applied with respect to the states $Y_\pm$ and so forth, completing the
proof. \qed

Note that the proof is constructive and allows to determine a measurement strategy
that would at least achieve the bound reported.

Let us now show that the above result can also be generalized to a
setup where the spins $i$ and $j$ can have any dimension, but the
measurements are still performed on qubits. To be more specific we
consider the  operator $S_A^i \otimes S_B^j$ acting on a bipartite
state $\rho_{ij}$ of arbitrary dimension. Since local unitary
transformations can always be absorbed in the definition of
$\rho_{ij}$ we can choose $S_A^i$ and $S_B^j$ to be diagonal. The
correlation function can then be written in the bilinear form:
{\small \bea Q_{AB}^{ij}&=&\tr [\rho_{ij} \ (S_A^i \otimes
S_B^j)]-\tr [\rho_{ij} \ (S_A^i \otimes \one) ]\tr[\rho_{ij} \ (\one \otimes S_B^j)] \nonumber \\
&=&\frac{1}{2} \vec{x}^T (\vec{a} \ \vec{1}^T- \vec{1} \ \vec{a}^T) \otimes (\vec{b} \ \vec{1}^T- \vec{1} \  \vec{b}^T)\vec{x}\ ,
\eea }
where the column vectors $\vec x, \vec a$
and $ \vec b$ are representing the diagonal elements of the matrices
$\rho_{ij}, S_A^i$ and $S_B^j$, and $\vec 1$ is a column vector with all ones.
Thus the matrix
$Z:=(\vec{a} \ \vec{1}^T-
\vec{1} \ \vec{a}^T) \otimes (\vec{b} \ \vec{1}^T- \vec{1} \ \vec{b}^T)$
replaces the
matrix $\sigma_y\otimes \sigma_y$ in the definition (\ref{Smatrix}). $Z$ is the tensor product
of two antisymmetric matrices of rank two and therefore has two positive and two
negative eigenvalues. This property is sufficient to fulfill the inequality
(\ref{ineq2}).

\section{Maximum correlation functions and entanglement of pure bipartite states}
\subsection{Two-qubit states}
For an arbitrary two-qubit state $\rho$ we want to maximize the correlation
function:
{\small \be \label{Qab}
 Q_{AB}^{ij}= \tr[\rho ( \ S^i_A \otimes S^j_B)]-\tr[\rho ( \
S^i_A \otimes \one)]\tr[\rho  (\one \otimes S^j_B)].
 \ee }
For qubits  we can parametrize $S_A$ and $S_B$ by the
three-dimensional unit vectors $\vec a, \vec b$: \bea
S_A&=& \vec \sigma \cdot \vec a  \ ,\\
S_B&=& \vec \sigma \cdot \vec b \ , \eea where $\vec \sigma=(\sigma_x \  \sigma_y
\  \sigma_z)$.
 The correlation can then be written in the form:
\be Q_{AB}=
\sum_{\alpha \beta}  a_\alpha \ Q_{ \alpha \beta} \ b_\beta =: \vec a^T\  Q \ \vec
b \quad \quad (\alpha, \beta=x, y, z) \ . \ee
The matrix elements $Q_{ \alpha \beta}$ of the $3 \times 3$ matrix $Q$ are defined by (\ref{Qab}) with $S_A=\sigma_\alpha$, $S_B=\sigma_\beta$. Clearly the maximum value for $Q_{AB}$ is given by the largest
singular value of the matrix $Q$.

 For a pure state $\rho=|\psi\ra \la \psi |$ the matrix $Q$ can be computed using the Schmidt decomposition $|\psi\ra=\sum_i \ \lambda_i \ |i_A \ra \otimes |i_B \ra $ with $\lambda_i \geq 0$. Note that local unitary transformations can always be absorbed in the definition of $S_A$ and  $S_B$. In this representation the matrix $Q$ is diagonal and one can show that the maximum value is given by $Q_{xx}=2 \lambda_1 \lambda_2$. It  can easily be checked that this expression is equal to the concurrence $C$ as defined in (\ref{DefC}). Thus we have shown that the entanglement of a pure two-qubit state
$|\psi\ra $ as measured by the concurrence $C$ is equal to the maximum correlation
function: \be \max_{\vec a , \vec b} (Q_{AB}(|\psi\ra))=Q_{xx}(|\psi\ra)=C(|\psi\ra)=2
\lambda_1 \lambda_2 \ . \ee This relation is central for establishing the strong
connection between classical and quantum correlations in pure multipartite qubit
states (see Sect. III).
\subsection{Two-qutrit states}
As in the qubit case we consider  correlations of the form (\ref{Qab}).  In generalization to the usual
spin-1 operators we want to maximize with respect to the  bounded operators $- \openone \leq S_A, S_B \leq \openone$. In the forthcoming discussion we will only
consider pure states. In Schmidt decomposition we have $|\psi\ra = \sum_{i=1}^3
\lambda_i \ |i\ra \otimes |i\ra$ with $\lambda_i \geq 0$ and
$\lambda_1^2+\lambda_2^2+\lambda_3^2=1$. Let us begin with rewritting the
correlation function: \bea
Q_{AB}&=& \tr_A(S_A \ (\rho_1 -\beta  D)  \label{Q1} \\
&=& \tr_B(S_B \ (\rho_2 -\alpha  D) \ . \eea Here we have defined the $3
\times 3$ matrices: \bea
\rho_1&:=&\tr_B({\bf 1} \otimes S_B \ \rho) = D^\frac{1}{2} S_B^T D^\frac{1}{2} \ ,\\
\rho_2&:=&\tr_A(S_A \otimes {\bf 1}  \ \rho) = D^\frac{1}{2} S_A D^\frac{1}{2} \ ,\\
D&:=& \rm{diag}(\lambda_1^2,  \lambda_2^2,  \lambda_3^2 ) \ , \eea and the scalars
$\alpha=\tr (\rho_2)$,  $\beta=\tr(\rho_1)$. We further introduce the
eigenvalue decomposition $\rho_1 -\beta  D=U E U^\dagger$. Note that  the diagonal
matrix $E$ has zero trace and thus  has at least one negative entry. Now  one
immediately sees from (\ref{Q1}) that $Q_{AB}$ is maximized if $S_A$ has the same
eigenvectors as $\rho_1 -\beta  D$ and if its eigenvalues are given by the sign of
the matrix $E$. Hence we can formulate the following relations that  hold for the
maximum correlation function: \bea
&(i)& S_A=U \ \rm{sign}(E) \ U^\dagger \quad \rm{ with } \ \ \tr( E) =0 \\
&(ii)& [S_A,  \rho_1 -\beta  D]=0 \\
&(iii)& [S_B,  \rho_2 -\alpha  D]=0 \ . \eea These conditions lead to the simple
commutator relation $[D,M]=0$ where $M:=U(|E| - \alpha \ E ) U^\dagger$. Since $M$
commutes with the diagonal matrix $D$ it has to be diagonal. Trivially this is
fulfilled for diagonal $U$ implying also a diagonal operator $S_A$. A nondiagonal
$U$ is only possible if the matrix $(|E| - \alpha \ E )$ is degenerate.  Hence we
arrived at the surprising result that the operator $S_A$ that maximizes the
correlation $Q_{AB}$ is either diagonal $S_A=\rm{diag}(1, -1, -1)$ or can be
parameterized in the form: \be \label{SA}
S_A=\left( \begin{array}{ccc} 1 & 0& 0 \\
0& \cos(\theta)& \sin(\theta) e^{-i \phi} \\
0&  \sin(\theta) e^{i \phi} & -\cos(\theta)    \end{array}  \right) \ . \ee

As $Q_{AB}$ is symmetric in A and B an equivalent expression with rotation angles
$\theta'$ and $\phi'$ holds for the operator $S_B$. From this we can deduce the
relations $\alpha= \lambda_1^2+ \cos(\theta') (\lambda_2^2-\lambda_3^2)$ and
$\beta= \lambda_1^2+ \cos(\theta) (\lambda_2^2-\lambda_3^2)$.

The required degeneracy of the matrix  $(|E| - \alpha \ E )$ puts a constraint on
$\alpha$ (or the optimal rotation angle $\theta'$) as  a function of $\beta$:
$\alpha =F(\beta)$ (for simplicity we do not specify the function $F$ here). Due to symmetry it
also holds $\beta=F(\alpha)$. Clearly a fixpoint of the maximization procedure is
given by the symmetric solution $\alpha=\beta$ (or $\theta =\theta'$). However,
there also exists an asymmetric solution. In order to obtain nice analytical
expressions for these solutions it is more convenient to parameterize the function
$Q_{AB}$ using the form (\ref{SA}) and then maximize with respect to the rotation
angles $\theta$ and $\theta'$: {\small \bea
\hspace*{-5mm} Q_{AB}&=& \lambda_1^2+ \cos\theta \cos\theta' (\lambda_2^2+\lambda_3^2) + 2 \sin{\theta} \sin \theta' \lambda_2 \lambda_3  \nonumber \\
&-&(\lambda_1^2+(\lambda_2^2-\lambda_3^2)\cos
\theta)(\lambda_1^2+(\lambda_2^2-\lambda_3^2)\cos \theta')  \ . \label{QABtheta}
\eea}
Here we made the choice  $\phi=-\phi'$, which maximizes $Q_{AB}$. We further
note that in this expression the role of the Schmidt coefficient $\lambda_1$ is
special, which results from the ordering $\lambda_1 \leq \lambda_2 \leq
\lambda_3$. For the symmetric case ($\theta=\theta'$) we obtain the optimal
rotation angle: \be \cos(\theta_{opt})= \frac{\lambda_1^2
(\lambda_2^2-\lambda_3^2)}{(\lambda_2-\lambda_3)^2-(\lambda_2^2-\lambda_3^2)^2} \
, \ee which yields the maximum correlation function: \be Q_{AB}^{sym}=\frac{4
\lambda_2^2  \lambda_3^2}{2 \lambda_2 \lambda_3-\lambda_1^2} \ . \ee Notice that
for $\lambda_1=0$ this reduces to the qubit solution $Q_{AB}^{max}=2 \lambda_2
\lambda_3$. As for the symmetric case $Q_{AB} $ (\ref{QABtheta}) is quadratic in
$\cos(\theta)$ the maximum can also be reached at the boundaries $\cos(\theta)=\pm
1$. This leads to  diagonal operatators $S_A=S_B$ and the  maximum correlation is
given by: \be Q_{AB}^{diag}=1-(\lambda_3^2-(\lambda_2^2+\lambda_1^2))^2 \ . \ee
The asymmetric solution can also be worked out, but is difficult to cast  in a
nice analytical form. For our purpose it is enough to establish  the following
relation for the optimal rotation angles: \bea
&a& (\cos(\theta) +\cos(\theta'))= -b (1+ \cos(\theta) \cos(\theta') \label{thetaas}\\
&a&:=\lambda_2^2+\lambda_3^2 -(\lambda_3^2-\lambda_2^2)^2\\
&b&:=\lambda_1^2(\lambda_3^2 -\lambda_2^2) \ . \eea It can easily be verified that
$a, b>0$ and $a \geq b$. Inserting (\ref{thetaas}) in $Q_{AB}$ (\ref{QABtheta}) it
follows that the asymmetric solution can be upper bounded by: \be Q_{AB}^{asym}
\leq  \lambda_1^2 -\lambda_1^4 - \frac{b^2}{a}+2 \lambda_2 \lambda_3 \ . \ee
Straightforward analysis shows that the two-qubit limit ($\lambda_1=0$) yields an
\emph{upper} bound for the maximum correlation in the  two-qutrit case: \be
\label{QABub} Q_{AB}^{max}:=\max(Q_{AB}^{sym},Q_{AB}^{diag}, Q_{AB}^{asym}) \leq 2
\lambda_2 \lambda_3|_{\lambda_1=0}. \ee The maximum correlation function
decreases if the number of non-zero Schmidt coefficients increases. Thus
$Q_{AB}^{max}$ cannot be used for measuring  entanglement as in the qubit case.
The entropy of entanglement $E(\ket{\psi})$ \cite{Bennett96}, on the contrary,
increases with the number of  non-zero Schmidt coefficients. A fact that follows
directly from the concavity property of $E(\ket{\psi})$. Hence the $\lambda_1=0$
case $E(\ket{\psi})=f(2 \lambda_2 \lambda_3|_{\lambda_1=0})$, with $f$ being the
convex function (\ref{convexf}), yields a \emph{lower} bound on the entropy of
entanglement. From this it follows that the entanglement of a pure two-qutrit
state is lower bounded by the maximum correlation function: \be E(\ket{\psi})\geq f(Q_{AB}^{max}) \ . \ee

\section{Analytical calculation of the string order parameter and the LE for matrix product states}
We consider a MPS (\ref{MPS}) with qubit bonds \mbox{($D=2$)}. In the case of OBC and qubits at the endpoints \mbox{($i=0,N+1$)} this (unnormalized) MPS state can be  written in the form:
\be \label{MPSAKLT}
| \psi \ra =\sum_{\alpha, i_1 \ldots i_N, \beta} \ {\vec{a}}^\alpha A^{i_1}\ldots A^{i_N} { \vec b}^\beta \ | \alpha\ra | i_1 \ldots i_N\ra |\beta \ra ,
\ee
where $ \vec a$ and $ \vec b$ are two dimensional row and column  vectors, respectively, and $\alpha,\beta \in \{0,1 \}$.  We are interested in the string order parameter (\ref{Qso}) between the endspins. Using expression (\ref{Exp_Val1}) for calculating expection values of MPS,  we can write:
 \be
 Q_{SO}^{0,N+1}=  \frac{\vec{E}^a_{\sigma_z} ({E}_{R})^N \vec{E}^b_{\sigma_z} }{{\vec E}^a_{\openone} {(E_{\openone})}^N {\vec E}^b_{\openone}}.
 \ee
In the limit of large $N$ and diagonalizable ${E}_{R}$ ($E_{\openone}$) only the maximum eigenvalue  $\lambda_R$ ($\lambda_{\openone}$) will survive:
\be
 \xi_{SO}:= \lim_{N\rightarrow \infty} Q_{SO}^{0,N+1}=   \frac{(\vec{E}^a_{\sigma_z}  {\vec r_R}) ({\vec l_R}  \vec{E}^b_{\sigma_z} )}{(\vec{E}^a_{\openone}  {{\vec r}_{\openone}}) ({\vec l_{\openone}}  \vec{E}^b_{\openone} )} \left( \frac{\lambda_R}{\lambda_{\openone}} \right)^N,
 \ee
where ${{\vec l}_{O}}$ and  ${\vec r_{O}}$  denote the left and right eigenvectors of $E_O$. \\
In the case of the  AKLT model and for  the basis (\ref{UHeis}), we have: $A^1 = i \sigma_y$, $A^2 =  \sigma_z$ and $A^3 =  \sigma_x$.  Hence one finds
\bea
E_{\openone}&=& \sigma_x \otimes \sigma_x - \sigma_y \otimes \sigma_y+  \sigma_z \otimes \sigma_z,  \label{E1}\\
E_{R}&=& -\sigma_x \otimes \sigma_x + \sigma_y \otimes \sigma_y+  \sigma_z \otimes \sigma_z, \label{Er}
\eea
and  $\lambda_R= \lambda_{\openone}=3$. Realizing that $ {\vec{a}}^\alpha $ and ${ \vec b}^\beta $ are representing unit vectors in the standard basis, we obtain the  result: $\xi_{SO}=1$.

 Let us now show how to calculate the LE between the end points of the chain for states of the form (\ref{MPSAKLT}).  Since the end spins are represented by qubits we can use the concurrence (\ref{DefC}) as entanglement measure, which simplifies the calculation considerably.  For the basis $\mathcal{M}=\{ |i\ra \la i | \}$ the average entanglement can be written as \cite{VMC04}:
 \be
 L^{\mathcal{M},C}_{0,N+1}=   \frac{\sum_{ i_1 \ldots i_N} 2  \ | \det( A^{i_1}\ldots A^{i_N} )|}{{\vec E}^a_{\openone} {(E_{\openone})}^N {\vec E}^b_{\openone}}.
 \ee
Since the determinant factorizes, we obtain
 \be \label{LE_MPS}
 L^{\mathcal{M},C}_{0,N+1}\rightarrow  \frac{2}{(\vec{E}^a_{\openone}  {{\vec r}_{\openone}}) ({\vec l_{\openone}}  \vec{E}^b_{\openone} )} \left( \frac{\sum_{ i}  | \det( A^{i})|}{\lambda_{\openone}} \right)^N,
 \ee
in the limit of large $N$. The basis which maximizes $ L^{\mathcal{M},C}$ is clearly the same basis, which maximizes the expression $\sum_{ i}  | \det( A^{i})|$. This problem is equivalent to calculating the EoA of the $D^2 \times D^2$ state $A^\dagger A$:
\be
E_A(A):= \sup_\mathcal{M} \sum_{ i}  | \det( A^{i})| = \tr|A^T(\sigma_y\otimes\sigma_y)A|.
\ee
 The elements of  the $(2S+1)\times D^2$ matrix  $A$ are given by $A_{i,(\alpha \beta)}=A^i_{\alpha, \beta}$.
 Hence we found a necessary and sufficient condition for long range order in the entanglement (i.e. non-vanishing $ L^C_{0,N+1}$ for $N\rightarrow \infty$): The expression $E_A(A)$ has to be equal to the largest eigenvalue, $\lambda_{\openone}$,  of the matrix $E_{\openone}$.
For the AKLT model, one can easily check that this condition is indeed fulfilled,  and that  $L_{0,N+1}^C=1$. \\
The ground state of the AKLT thus exhibits long range order both in terms of the  LE and the string order parameter.

\end{appendix}


\begin{thebibliography}{10}
\bibitem{comLOCC} For pure four qubit states it has been shown in \cite{V01}, that in the case of  projective measurements the LE cannot be increased by classical communication. In addition, numerical studies for the same system indicated that POVM measurements lead only to a very small   increase  of LE, compared to projective measurements.
\bibitem{comMC} Note that with  this definition  the number of
sweeps $M$ is independent of the system size $N$.
\bibitem{comMPS} We  note that for  antiferromagnetic systems only every second matrix
$U^{[k],s}$ converges, resulting in an alternating MPS structure
$A^{s_1}B^{s_2}A^{s_3}\ldots B^{s_N}$. One can show that the translationally
invariant state which minimizes the energy is then given by the linear
superposition  $A^{s_1}B^{s_2}A^{s_3}\ldots B^{s_N}+ B^{s_1}A^{s_2}B^{s_3}\ldots
A^{s_N}$.
\bibitem{comMPS1} As a first test of our method we calculate the energy per site of an  infinite chain, using the translationally invariant MPS.  We confirmed that the result is as accurate as the one obtained directly from   DMRG.
\bibitem{comfluc} We note that for $\lambda < 1$ the ratio $\delta L_{i,j}^C / L_{i,j}^C$ can be much larger than one.
\bibitem{comAKLT} We note that the degeneracy in the case of OBC could be lifted by placing $S=1/2$ spins at the ends. For technical reasons we do not consider this situation.
\bibitem{com1} In the case of a singular matrix $R$ Sylvester's
law of inertia cannot directly be applied. However, it is guaranteed that in the
worst case the principal block has  one zero eigenvalue, which would lead to the
equality with the original correlations.
\end{thebibliography}
\end{document}